\documentclass[reprint, amsmath,amssymb, aps, ]{revtex4-2}

\usepackage[utf8]{inputenc} 
\usepackage{graphicx}
\usepackage{dcolumn}
\usepackage{bm}
\usepackage{amsmath}
\usepackage{enumitem}
\usepackage{hyperref} 
\usepackage[user]{zref}

\hypersetup{
	colorlinks=true,
	linkcolor=blue,
	citecolor=blue,
	filecolor=blue,
	urlcolor=blue,
} 

\newcommand{\ve}[1]{\mathbf{#1}}
\newcommand{\te}[1]{\overline{\overline{#1}}}

\newcommand{\figref}[2]{\hyperref[#1]{Fig. \ref*{#1}{#2}}}
\newcommand{\secref}[1]{Sec.~\ref{#1}}
\newcommand{\subsecref}[1]{Sec.~\ref{#1}}

\newcommand{\eqqref}[1]{\hyperref[#1]{Eq.~\eqref{#1}}}

\def\@#1{_{\rm #1}}

\begin{document}

\preprint{APS/123-QED}

\title{Bound States in the Continuum as Converging Scattering Zeros in Metasurfaces:\\ Why Accidental BICs Require Nonlocality}


\author{Karim Achouri}
\email{karim.achouri@epfl.ch}
\author{Youssef Dib}
\author{Hossein Allahverdizadeh}

\affiliation{Institute of Electrical and Microengineering, École Polytechnique Fédérale de Lausanne, Laboratory for Advanced Electromagnetics and Photonics, Lausanne, Switzerland}

\author{Sergejs Boroviks}
\affiliation{Nanophotonics and Metrology Laboratory, Swiss Federal Institute of Technology Lausanne (EPFL), Lausanne, Switzerland}

\author{Mahmoud Elsawy}
\affiliation{Université Côte d'Azur, Inria, CNRS, LJAD, Sophia Antipolis, France}

\date{\today}

\begin{abstract}

Bound states in the continuum (BICs) allow metasurfaces to achieve extremely high quality factors, yet it remains unclear how they emerge and how they can be described by effective material parameters. We address this question by modeling BICs through the reflection and transmission zeros of a metasurface described by effective susceptibilities. By decomposing its response into even and odd parts, and inspecting the interaction of the corresponding poles and zeros, we show that a high-Q resonance appears as a transmission zero squeezed between two reflection zeros, or vice versa. The separation between these scattering zeros is set by the linewidth of the corresponding even or odd resonance, so that a BIC corresponds to a convergence of these zeros. This simple concept explains symmetry-protected BICs and their transformation into quasi-BICs. However, it also reveals a fundamental limitation for accidental BICs: in a passive local dipolar model, the zeros cannot converge at oblique incidence unless one of the susceptibility resonances has a negative amplitude, which would make the metasurface active according to Foster's theorem. We demonstrate that this negative amplitude is in fact the signature of higher-order multipoles or lattice coupling effects that the local model cannot properly capture, which shows that accidental BICs are intrinsically nonlocal. We confirm this conclusion by simulating a silicon-ridge metasurface, whose response is reproduced and predicted by a passive, spatially dispersive model, which also describes how breaking the mirror symmetry of the metasurface turns the accidental BIC into a quasi-BIC.

\end{abstract}


\maketitle

\section{Introduction}
\label{sec_intro}

Bound states in the continuum (BICs) can be defined as the zero-linewidth limit of resonances that do not radiate into free space even though their frequency lies within the radiation continuum~\cite{hsu2016}. While true BICs are hardly achievable in practice, quasi-BICs are ubiquitously used in periodic nanostructures to achieve high quality factor (Q) spectral features~\cite{hsu2013,koshelev2018,jin2019,koshelev2023,schulz2024,yu2025}, for instance in asymmetric waveguide gratings~\cite{quaranta2018,yezekyan2024}, in strongly coupled resonant structures~\cite{allayarov2025a,bogdanov2019}, and in dielectric~\cite{huang2023} and plasmonic metasurfaces~\cite{liang2026}.

BICs are usually separated into symmetry-protected BICs, which typically occur at normal incidence due to symmetry mismatch, and accidental BICs, which occur at a specific oblique incidence angle and wavelength as a result of parameter tuning~\cite{friedrich1985,hsu2016}. Breaking the symmetry that protects a BIC, for instance with a substrate or structural imperfections, turns it into a leaky resonance with finite Q, i.e., a quasi-BIC~\cite{sadrieva2017}. Since a BIC requires an exact cancellation of the far-field radiation, it often arises from destructive interference between several radiating multipoles rather than from the vanishing of a single one~\cite{sadrieva2019,gladyshev2022}. Changing the incidence angle can break the symmetry or detune this interference, which makes BICs strongly angle-dependent and can be leveraged in angle-selective metasurfaces~\cite{arslan2017,achouri2020a,zhang2020}. Note also that the concept of BIC has been extended to non-Hermitian systems~\cite{Sakotic2023,binkowski2024,valero2025}, which is relevant here since we find that a purely dipolar model of an accidental BIC can require effective parameters with a gain-like response.

Such angle-dependent responses are conveniently described with effective material parameters~\cite{tretyakov2003,kuester2003,idemen2011,achouri2015,epstein2016}. Here, we use effective surface susceptibilities connected to the incident and scattered fields via generalized sheet transition conditions (GSTC)~\cite{achouri2021}. Originally, this approach was limited to dipolar susceptibilities that do not depend on the incidence angle, but it has now been extended either by accounting for higher-order multipoles with angle-independent susceptibilities~\cite{achouri2022b,tiukuvaara2023,allahverdizadeh2025a}, or by allowing the susceptibilities to depend on the incidence angle, i.e., to be spatially dispersive~\cite{rahmeier2022a,rahmeier2023}. These two approaches lead to similar responses, and mapping relations have been developed to connect them~\cite{dugan2022a}. Such spatially dispersive, or nonlocal, responses are also at the core of nonlocal metasurfaces, which often exploit quasi-BICs~\cite{kwon2018,overvig2020a,overvig2022,shastri2023}. To avoid ambiguity, in this work, we consider any response to be nonlocal if it involves bianisotropy, which corresponds to weak spatial dispersion, spatially dispersive (angle-dependent) susceptibilities, higher-order multipoles or lattice coupling interactions. As we will see, spatially dispersive susceptibilities, which may result from higher-order multipoles or lattice coupling, are necessary to describe accidental BICs using effective material parameters, while bianisotropy is needed to model the transition of an accidental BIC into a quasi-BIC~\cite{asadchy2016}.

\begin{figure*}[t]
\centering
\includegraphics[width=\linewidth]{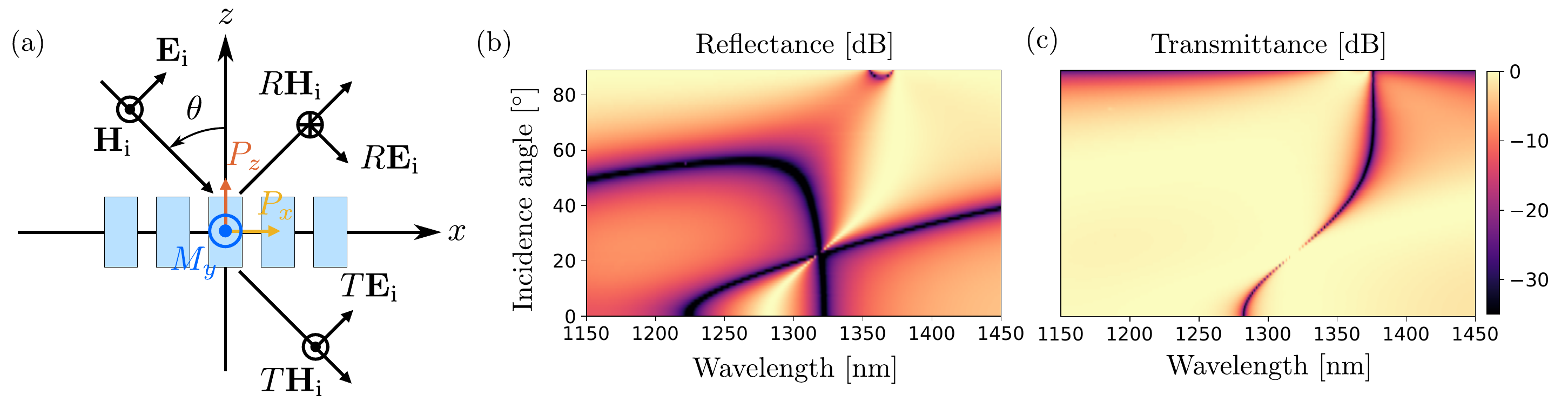}
\caption{TM-polarized scattering from a metasurface made of a one-dimensional periodic array of silicon ridges ($\epsilon_\text{r}=12$) with a width of 280~nm and a height of 740~nm. The array period is 500~nm. The background medium is vacuum. (a) The interactions between the waves and the metasurface excite the dipolar polarization densities $M_{y}$, $P_{x}$ and $P_{z}$. (b) Numerically simulated reflectance and (c) transmittance of the metasurface.}
\label{fig_BIC}
\end{figure*}

In this work, we use the GSTC framework to study the emergence of BICs in terms of the scattering zeros of a metasurface, i.e., the combinations of susceptibilities and incidence angle for which the reflected or transmitted field vanishes, independently of the geometry and material of the structure. This perspective reveals that real reflection and transmission zeros appear near BICs in symmetric periodic structures~\cite{shipman2012,yuan2022,Sakotic}, where a BIC corresponds to a pole and a zero of the scattering matrix that merge on the real frequency axis~\cite{blanchard2016,bogdanov2019,Krasnok2019a}. While it has recently been shown that reflection zeros can be used to derive susceptibility conditions for invisibility~\cite{yucel2026}, and quasi-BICs have been connected to effective material parameters~\cite{abdipour2026}, the intricate relation between BICs and scattering zeros has not been shown so far. A more conventional approach for derivation of analytical conditions for accidental BICs~\cite{abujetas2020,abujetas2022,Bykov2019} and Brewster-like quasi-BICs~\cite{abujetas2019} uses the polarizabilities of individual particles combined with lattice sums. While such a microscopic approach is interesting because it provides a direct access to the coupling between the particles, it comes at the cost of non-analytical lattice sum computations that are needed to relate the scattered fields to the polarizabilities. This makes the analysis of a particular response, such as a scattering zero, significantly more complicated. In contrast, the effective susceptibilities of the GSTC are directly related to the reflected and transmitted fields, since they already include the coupling between the particles~\cite{achouri2015,achouri2021}. This makes our approach particularly convenient to investigate specific scattering responses, such as the scattering zeros associated with BICs, at the cost of not being able to access lattice coupling effects.

This work is structured as follows. In~\secref{sec_model}, we start by modeling a metasurface using the GSTC framework and distinguishing its even and odd scattering coefficients, whose poles correspond to the resonant modes of the metasurface. In~\secref{sec_zeros}, we derive closed-form conditions for the reflection zeros, invisibility, and transmission zeros in terms of the effective susceptibilities and the angle of incidence. We also relate the separation between these zeros to the linewidth of the resonances and to the poles and zeros of the scattering matrix. To relate these zeros to the actual resonances of scattering particles, we then express the susceptibilities with Lorentzian functions so as to mimic their spectral response. In~\secref{sec_examples}, we use this Lorentzian model to investigate how the resonances of the different susceptibility components lead to symmetry-protected BICs, quasi-BICs, avoided crossings of the scattering zeros, and accidental BICs. We then compare our analytical model with the full-wave simulation and multipole analysis of the response of a metasurface made of silicon ridges. This comparison shows that a passive accidental BIC cannot be described by a local dipolar model, which motivates us to introduce in~\secref{sec_passive} a passive spatially dispersive model whose parameters are extracted from an eigenmode analysis of the silicon-ridge metasurface. This also allows us to show how breaking the mirror symmetry $z\rightarrow -z$ transforms the accidental BIC into a quasi-BIC. We finally conclude our discussion in~\secref{sec_conclusion}. The derivations along with additional examples are provided in the appendices.

\section{Metasurface Model and Even--Odd Decomposition}
\label{sec_model} 

We consider a metasurface to be a subwavelength array composed of a periodic arrangement of scattering particles. We assume the array period to be small enough compared to the wavelength so that only zeroth-order diffraction can take place. In that case, a metasurface may be modeled as a zero-thickness sheet of effective electric and magnetic current densities using the GSTC~\cite{tretyakov2003,kuester2003,achouri2015,epstein2016,idemen2011}. These currents may generally be expressed in terms of multipole moments that are themselves proportional to effective susceptibilities~\cite{achouri2021,achouri2022b,tiukuvaara2023,allahverdizadeh2025a}. For a metasurface lying in the $xy$-plane at $z=0$, the dipolar form of the GSTC may be expressed as~\cite{achouri2021} 
\begin{subequations}
\label{eq_GSTC} 
\begin{align}
\ve{\hat{z}}\times\Delta\ve{H} &= +j\omega\ve{P}_\parallel - \frac{1}{\mu_0}\ve{\hat{z}}\times\nabla_\parallel M_z,\\
\ve{\hat{z}}\times\Delta\ve{E} &= -j\omega\ve{M}_\parallel - \frac{1}{\epsilon_0}\ve{\hat{z}}\times\nabla_\parallel P_z,
\end{align}
\end{subequations}
where $\Delta\ve{E}$ and $\Delta\ve{H}$ denote the field differences at the metasurface, and $\ve{P}$ and $\ve{M}$ are the surface polarization densities excited on the metasurface.

To simplify the discussion, we start by considering an anisotropic metasurface described only by the diagonal components of the susceptibility tensors $\te{\chi}_\text{ee}$ and $\te{\chi}_\text{mm}$, and illuminated by a TM-polarized plane wave propagating in the $xz$-plane. The case of TE polarization is not discussed here for conciseness but may be treated in a similar way. Such a uniaxial or biaxial structure has no bianisotropic or gyrotropic response, so that only the polarizations $M_{y}$, $P_{x}$ and $P_{z}$ can be excited, as illustrated in Fig.~\ref{fig_BIC}a for an array of spatially symmetric dielectric resonators in a homogeneous background medium, whose reflectance and transmittance are plotted in Figs.~\ref{fig_BIC}b and~\ref{fig_BIC}c. Note that the following developments are, however, not limited to dielectric resonators.

Following~\cite{achouri2021}, we express these polarizations in terms of effective surface susceptibilities as
\begin{subequations} 
\label{eq_const} 
\begin{align} 
M_y &= \mu_0 \chi_\text{mm}^{yy} H_{y,\text{av}},\\
P_x &= \epsilon_0 \chi_\text{ee}^{xx} E_{x,\text{av}},\\
P_z &= \epsilon_0 \chi_\text{ee}^{zz} E_{z,\text{av}}\label{eq_const_pz},
\end{align} 
\end{subequations}
where $\ve{E}_\text{av}$ and $\ve{H}_\text{av}$ denote the field averages at the metasurface.

Substituting~\eqref{eq_const} into~\eqref{eq_GSTC} and solving for the TM-polarized reflection and transmission coefficients leads to~\cite{achouri2020a} 
\begin{subequations}
\label{eq_RT} 
\begin{align}
    R(\theta) &= 2j\frac{k^2(\chi_\text{ee}^{xx}-\chi_\text{mm}^{yy})-k_x^2(\chi_\text{ee}^{xx}+\chi_\text{ee}^{zz})}{(2+jk_z\chi_\text{ee}^{xx})\left[2k_z+j(k^2\chi_\text{mm}^{yy}+k_x^2\chi_\text{ee}^{zz})\right]},\\
    T(\theta) &= \frac{k_z\left[4+(k^2 \chi_\text{mm}^{yy}+k_x^2 \chi_\text{ee}^{zz})\chi_\text{ee}^{xx}\right]}{(2+jk_z\chi_\text{ee}^{xx})\left[2k_z+j(k^2\chi_\text{mm}^{yy}+k_x^2\chi_\text{ee}^{zz})\right]}\label{eq_RTt},
\end{align}
\end{subequations}
where $k$ is the wavenumber in the background medium, here vacuum, $k_{x}= k \sin \theta$ and $k_{z}= k \cos \theta$.

Since the metasurface is symmetric with respect to the plane $z=0$, its response can be split into two independent parts. Under the mirror transformation $z\rightarrow -z$, the polarization $P_x$ remains unchanged and thus forms the \emph{even} part of the response, while $M_y$ and $P_z$ change sign and form the \emph{odd} part. Accordingly,~\eqref{eq_RT} may be rewritten as
\begin{equation}
\label{eq_RTeo}
R = \frac{s_\text{o}-s_\text{e}}{2},\qquad T = \frac{s_\text{o}+s_\text{e}}{2},
\end{equation}
with
\begin{subequations}
\label{eq_seso}
\begin{align}
s_\text{e} &= \frac{2-jk_z\chi_\text{ee}^{xx}}{2+jk_z\chi_\text{ee}^{xx}},\\
s_\text{o} &= \frac{2k_z-j(k^2\chi_\text{mm}^{yy}+k_x^2\chi_\text{ee}^{zz})}{2k_z+j(k^2\chi_\text{mm}^{yy}+k_x^2\chi_\text{ee}^{zz})},
\end{align}
\end{subequations}
where $s_\text{e}$ only depends on the even part of the response and $s_\text{o}$ only depends on the odd part. Physically, $s_\text{o}$ and $s_\text{e}$ describe how the metasurface scatters two symmetric illuminations made of two waves incident from both sides with the same or opposite magnetic fields at the metasurface, respectively. As shown in Appendix~\ref{app_S}, $s_\text{o}$ and $-s_\text{e}$ are the eigenvalues of the scattering matrix of the metasurface. The response to a single incident wave is thus simply the superposition of these two relations, which corresponds to~\eqref{eq_RT} via~\eqref{eq_RTeo}.

For a lossless metasurface, $|s_\text{e}|=|s_\text{o}|=1$. The scattering parameters in~\eqref{eq_RTeo} are thus entirely determined by the phase difference between $s_\text{e}$ and $s_\text{o}$. Moreover, the resonances of the metasurface correspond to the poles of $s_\text{e}$ and $s_\text{o}$ and each one is associated with a quasi-normal, i.e., complex eigenfrequency~\cite{lalanne2018}, mode of the metasurface, which, for the full-wave simulations of this work, we compute using the eigenfrequency solver of COMSOL (see related discussion in Sec.~\ref{sec_passive}). Solving for these poles gives the complex resonance frequencies of the metasurface, which we denote by $\tilde{\omega}$ to distinguish them from the real frequency $\omega$ of the incident wave. We write this complex frequency as $\tilde{\omega}=\omega_\text{r}+j\Gamma$, with $\omega_\text{r}=\text{Re}\{\tilde{\omega}\}$ and $\Gamma=\text{Im}\{\tilde{\omega}\}$, so that the Q factor of the corresponding resonance is $Q=\omega_\text{r}/(2\Gamma)$, where $2\Gamma$ is the linewidth of the resonance. Note that we use the time convention $e^{j\omega t}$, so that a resonance is damped when $\Gamma>0$. Finally, since $s_\text{e}$ and $s_\text{o}$ are independent of each other, the even and odd resonances are not coupled together, which is no longer the case when the mirror symmetry $z\rightarrow -z$ is broken, as discussed in~\subsecref{sec_zbreak}.

\begin{figure*}[t]
	\centering
	\includegraphics[width=\linewidth]{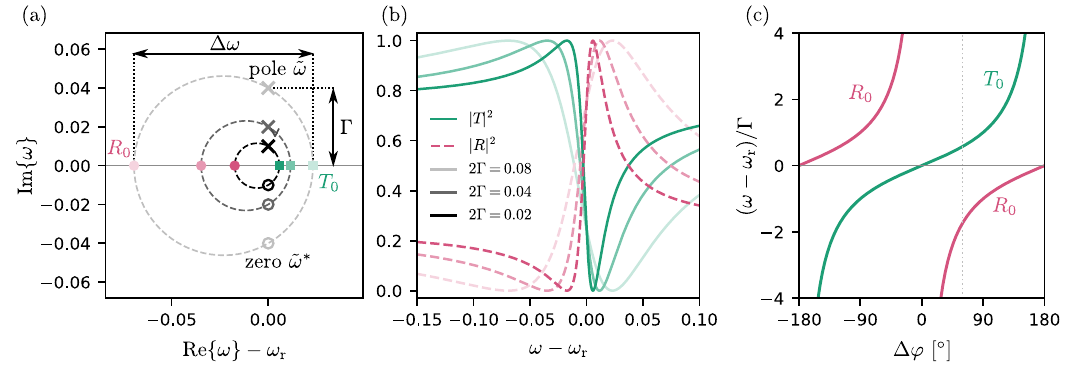}
	\caption{Scattering zeros near a resonance of $s_\text{o}$, computed with the single-resonance model of Appendix~\ref{app_zeros}, with frequencies normalized by $\omega_\text{r}$. (a) Pole $\tilde{\omega}$ ($\times$) and zero $\tilde{\omega}^*$ ($\circ$) in the complex frequency plane for complex values of $\omega$ and for the linewidths $2\Gamma=0.08$, $0.04$ and $0.02$ (light to dark), with $\Delta\varphi=60^\circ$. The $R_0$ (pink) and $T_0$ (green) conditions lie on the real axis, on the circle (dashed) passing through the pole and the zero. (b)~Corresponding transmittance (solid) and reflectance (dashed). (c)~Positions of the $R_0$ and $T_0$ conditions, normalized by $\Gamma$, versus $\Delta\varphi$. The lineshape evolves from a symmetric dip in transmission for $\Delta\varphi=0$ to a symmetric peak for $\Delta\varphi=\pm180^\circ$, through asymmetric Fano lineshapes. The dotted line indicates the value of $\Delta\varphi$ used in (a) and (b).}
	\label{fig_zeros}
\end{figure*}

\section{Scattering Zeros, Invisibility, and Resonances}
\label{sec_zeros}

The reflectance and transmittance maps in Fig.~\ref{fig_BIC} exhibit an accidental BIC at a wavelength of approximately 1320~nm and an incidence angle of $23^\circ$, which appears to emerge from the interplay of scattering zeros. Specifically, the BIC and the high-Q resonance in its vicinity seem to result from a transmission zero being ``squeezed'' by reflection zeros. To understand this mechanism, we now investigate how the metasurface scattering zeros may be expressed in terms of its susceptibilities. 

For a lossless metasurface, we consider three situations: 1) $R=0$ and $|T|=1$, i.e., a \emph{reflection zero}, denoted $R_{0}$, 2) $R=0$ and $T=1$, i.e., \emph{invisibility}, denoted $R_0^\text{inv}$, and 3) $|R|=1$ and $T=0$, i.e., a \emph{transmission zero}, denoted $T_{0}$. From~\eqref{eq_RTeo} and~\eqref{eq_RT}, these conditions are respectively given by
\begin{subequations}
\label{eq_S0} 
\begin{alignat}{2}
    R_{0}:&\; s_\text{o}=s_\text{e}, &\enspace &\sin^2\theta = \frac{\chi_\text{ee}^{xx}-\chi_\text{mm}^{yy}}{\chi_\text{ee}^{xx}+\chi_\text{ee}^{zz}},\label{eq_R0}\\
    R_0^\text{inv}:&\; s_\text{o}=s_\text{e}=1, &\enspace &\sin^2\theta = -\frac{\chi_\text{mm}^{yy}}{\chi_\text{ee}^{zz}},\label{eq_R0p}\\  
    T_{0}:&\; s_\text{o}=-s_\text{e}, &\enspace &\sin^2\theta = -\frac{4+k^2\chi_\text{ee}^{xx}\chi_\text{mm}^{yy}}{k^2\chi_\text{ee}^{xx}\chi_\text{ee}^{zz}}.\label{eq_T0}  
\end{alignat}
\end{subequations}
These conditions depend on both frequency and angle, but keep in mind that the susceptibilities are, for now, assumed not to depend on the angle. Note that at oblique incidence, the $R_0$ condition essentially corresponds to a generalized Brewster effect, as discussed in~\cite{paniagua-dominguez2016,lavigne2021,tiukuvaara2023}.

The invisibility condition~\eqref{eq_R0p} is the special case of~\eqref{eq_R0} for $s_\text{e}=1$, which, from~\eqref{eq_seso}, requires $\chi_\text{ee}^{xx}=0$. Invisibility is thus only possible if $\chi_\text{ee}^{xx}=0$, in which case~\eqref{eq_R0} reduces to the second expression in~\eqref{eq_R0p}.
In contrast, the $R_{0}$ and $T_{0}$ conditions are mutually exclusive because the metasurface is passive and lossless. They thus cannot be satisfied at the same wavelength and angle, since the incident power would then be neither reflected, nor transmitted, nor absorbed, which
would violate energy conservation. This also follows from~\eqref{eq_R0} and~\eqref{eq_T0}, since $s_\text{o}=s_\text{e}$ and $s_\text{o}=-s_\text{e}$ cannot be satisfied simultaneously.

Let us now investigate how these conditions behave near a resonance of $s_\text{o}$ with complex frequency $\tilde{\omega}$, while $s_\text{e}$ varies slowly. As shown in Appendix~\ref{app_zeros}, the phase of $s_\text{o}$ then changes by $2\pi$ over the resonance, which produces one $R_0$ and one $T_0$ condition on the real frequency axis separated in frequency by
\begin{equation}
\label{eq_dw_RT}
\Delta\omega=\frac{2\Gamma}{|\sin\Delta\varphi|},
\end{equation}
where $\Delta\varphi=\varphi_\text{bg}-\varphi_\text{e}$ is the phase difference between $s_\text{e}=e^{j\varphi_\text{e}}$ and the background $s_\text{bg}=e^{j\varphi_\text{bg}}$, i.e., the value of $s_\text{o}$ away from the resonance.

In terms of the scattering matrix, this resonance corresponds to a pole of $s_\text{o}$ at $\tilde{\omega}$ and, since the metasurface is passive and lossless, there must be a corresponding zero at $\tilde{\omega}^*$. The pole and the zero are thus separated by $2\Gamma$, which is directly proportional to the separation in~\eqref{eq_dw_RT}. As shown in Fig.~\ref{fig_zeros}a, the pole, the zero, and the $R_0$ and $T_0$ conditions all lie on a circle centered on the real axis, which contracts towards $\omega_\text{r}$ as the linewidth decreases~\cite{blanchard2016,kang2021}. The $R_0$ and $T_0$ conditions thus get closer as the linewidth decreases, which leads to an asymmetric Fano lineshape,~\cite{limonov2017} as shown in Fig.~\ref{fig_zeros}b, and ultimately merge together when it vanishes, which precisely corresponds to a BIC condition~\cite{bogdanov2019}. 

The phase shift $\Delta\varphi$ defines which of the zeroes is squeezed: if the resonance occurs close to an $R_0$ condition of the background, i.e., for $\Delta\varphi\approx 0$, the $T_0$ condition lies at the center of the resonance, while the $R_0$ condition is pushed away from it, on one side or the other depending on the sign of $\Delta\varphi$, as shown in Fig.~\ref{fig_zeros}c. Since $\Delta\varphi$ changes sign across the $R_0$ condition of the background, the $T_0$ condition then appears squeezed between two $R_0$ conditions leading to a resonance that appears as a narrow transmission dip. The accidental BIC in Fig.~\ref{fig_BIC} clearly represents this case.  On the other hand, for $\Delta\varphi\approx\pi$, an $R_0$ condition is squeezed between two $T_0$ conditions, and the resonance appears as a narrow transmission peak. For other values of $\Delta\varphi$, the resonance has an asymmetric Fano lineshape, as given in~\eqref{eq_fano} in Appendix~\ref{app_zeros}.

\section{Mechanisms for Bound States in the Continuum}
\label{sec_examples}

To show how a resonant behavior discussed in the previous section may be modeled in terms of susceptibilities, we next express them either with a single Lorentzian resonance or with a sum of several Lorentzian resonances of the form
\begin{equation} 
\label{eq_lorentz} 
    \chi_\text{aa}^{bb}(\omega) = \frac{A_\text{aa}^{bb}}{\omega_0^2 - \omega^2 + j\omega\gamma},
\end{equation}
where $\text{a}=\{\text{e,m}\}$, $b = \{x,y,z\}$, $A_\text{aa}^{bb}$ is an amplitude expressed in m$\cdot$s$^{-2}$, since surface susceptibilities are expressed in meters, $\omega_{0}$ is the resonance frequency, corresponding to the wavelength $\lambda_0=2\pi c/\omega_0$, and $\gamma$ is the damping factor, not to be confused with $\Gamma$. When several susceptibilities resonate, we denote the resonance wavelength of $\chi_\text{aa}^{bb}$ by $\lambda_\text{aa}^{bb}$, with the same indices as its amplitude $A_\text{aa}^{bb}$. Next, by deliberately setting $\gamma=0$, we restrict our attention to lossless metasurfaces  and consider several representative examples.

\subsection{Symmetry-Protected BIC}
\label{sec_MyPz}

We first consider a metasurface with $\chi_\text{ee}^{xx}=0$, i.e., exhibiting only $M_{y}$ and $P_{z}$, assuming that $\chi_\text{ee}^{xx}$ can be neglected within a considered wavelength and angular range. In that case, $s_\text{e}=1$, so that the response is fully determined by $s_\text{o}$ and any $R_0$ condition must therefore be equal to an invisibility condition $R_0^\text{inv}$.

\begin{figure}[b]
\centering
\includegraphics[width=\linewidth]{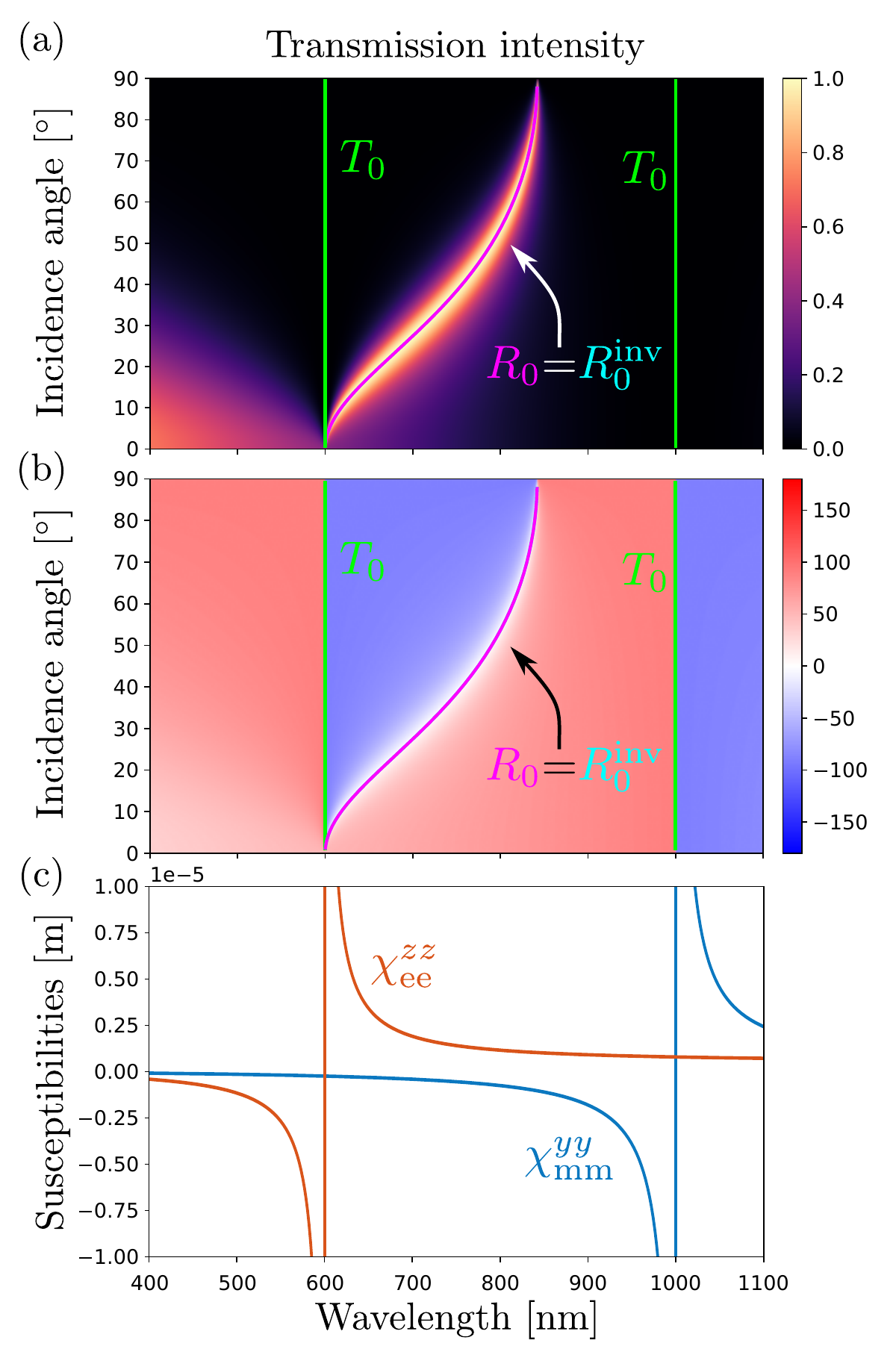}
\caption{Angular scattering response of a metasurface with $\chi_\text{ee}^{xx}=0$ and exhibiting only $M_{y}$ and $P_{z}$. The parameters of the Lorentzian resonances for the susceptibilities $\chi_\text{mm}^{yy}$ and $\chi_\text{ee}^{zz}$ are $A_\text{mm}^{yy}=1.5\times 10^{24}$, $A_\text{ee}^{zz}=5\times 10^{24}$, $\lambda_\text{mm}^{yy}=1000$~nm and $\lambda_\text{ee}^{zz}=600$~nm. (a) Transmittance, (b) transmission phase, and (c) real part of the susceptibilities. The pink and green solid lines denote the $R_{0}$ and $T_{0}$ conditions, respectively.}
\label{fig_PM_AmpPhase}
\end{figure}

Figure~\ref{fig_PM_AmpPhase} first shows an example of two angle-invariant $T_0$ conditions that are due to the resonances of $\chi_\text{mm}^{yy}$ and $\chi_\text{ee}^{zz}$, where $s_\text{o}\rightarrow -1$ leading to $R\rightarrow -1$ and $T\rightarrow 0$ for any angle.
We express $\chi_\text{mm}^{yy}$ and $\chi_\text{ee}^{zz}$ as single Lorentzian resonances with resonance wavelengths at $\lambda_\text{mm}^{yy}=1000$~nm and $\lambda_\text{ee}^{zz}=600$~nm, respectively, with the amplitudes given in the caption of Fig.~\ref{fig_PM_AmpPhase}, which shows the resulting transmission amplitude and phase versus wavelength and angle, together with the real parts of the susceptibilities. The $R_{0}$ and $T_{0}$ conditions are shown as solid pink and green lines, respectively.

The spectrum in Fig.~\ref{fig_PM_AmpPhase}a also shows a sharp dispersive transmission maximum feature across the full angular range. In the center of this feature lies an $R_0^\text{inv}$ condition, for which full transmission is achieved with zero phase shift, as can be verified in the phase map of Fig.~\ref{fig_PM_AmpPhase}b. This feature only exists between the two resonances, where $\chi_\text{mm}^{yy}$ and $\chi_\text{ee}^{zz}$ have opposite signs, so that~\eqref{eq_R0p} corresponds to a real angle. Since $s_\text{e}=1$ and the background is mostly reflective there, we have that $s_\text{o}\approx -1$ and $\Delta\varphi\approx\pm\pi$, which exactly corresponds to the second case discussed in~\secref{sec_zeros}. Near $\lambda=842$~nm, the feature vanishes into a high-Q tip, where $\chi_\text{mm}^{yy}=-\chi_\text{ee}^{zz}$, i.e., $\theta=90^\circ$ in~\eqref{eq_R0p}. 

At its other end, at $\theta=0^\circ$ and at the resonance of $\chi_\text{ee}^{zz}$, the feature also vanishes with a diverging Q factor. There, the $R_0$ and $T_0$ conditions converge, since all the conditions in~\eqref{eq_S0} tend to $\theta\rightarrow 0$ when $\chi_\text{ee}^{zz}\rightarrow\infty$. However, they only converge in a limiting sense, without ever crossing, which would violate energy conservation. What happens is that, at the point of convergence,~\eqref{eq_S0} breaks down and the scattering parameters reduce to 
\begin{subequations}
\begin{align}
    R(0) &= \frac{k\chi_\text{mm}^{yy}}{2j-k\chi_\text{mm}^{yy}},\\
    T(0) &= \frac{2j}{2j-k\chi_\text{mm}^{yy}},
\end{align}
\end{subequations}
which no longer depend on $\chi_\text{ee}^{zz}$ and differ from $R\rightarrow -1$ and $T\rightarrow 0$, obtained at the resonance of $\chi_\text{ee}^{zz}$ for $\theta\neq 0$. The scattering parameters are thus discontinuous at this point, which is a known behavior of BICs~\cite{shipman2012}. Convergence of the $R_0$ and $T_0$ conditions, with a diverging but inaccessible Q factor at $\theta=0^\circ$, is precisely the signature of a symmetry-protected BIC.

\begin{figure*}[!tp]
	\centering
	\includegraphics[width=0.85\linewidth]{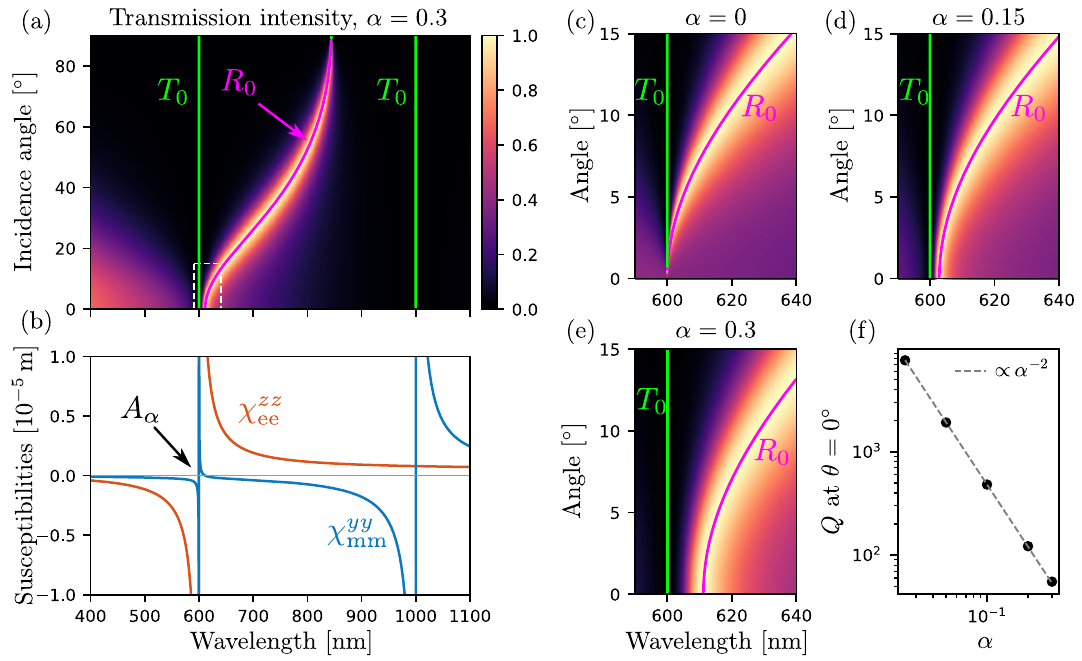}
	\caption{Quasi-BIC induced by an in-plane asymmetry of the particles. The metasurface is the same as in Fig.~\ref{fig_PM_AmpPhase}, with an additional Lorentzian resonance in $\chi_\text{mm}^{yy}$ at $\lambda=600$~nm with the amplitude $A_\text{mm}^{yy,\alpha}=A_0\alpha^2$, where $\alpha$ is the asymmetry parameter and $A_0=10^{24}$. (a) Transmittance for $\alpha=0.3$. The pink and green solid lines denote the $R_0$ and $T_0$ conditions, respectively. (b) Real part of the susceptibilities for $\alpha=0.3$. (c)--(e) Transmittance near the tip of the sharp transmission feature (dashed box in (a)) for $\alpha=0$ (BIC), and for $\alpha=0.15$ and $0.3$ (quasi-BICs), with the same color scale as in (a). (f) Q factor at normal incidence versus $\alpha$, obtained from the pole of $s_\text{o}$. The dashed line indicates the $\alpha^{-2}$ dependence.}
	\label{fig_qbic}
\end{figure*}

The divergence of $\chi_\text{ee}^{zz}$ leads to a BIC because, at normal incidence, the incident wave has no $z$-oriented electric field and therefore cannot excite $P_z$. In terms of symmetry, $P_z$ is the only dipolar polarization that does not change sign under the mirror transformation $x\rightarrow -x$, unlike the incident TM wave at normal incidence, which thus cannot couple to this mode.

We now derive an expression of the Q factor of this response by considering that the imaginary part of the resonance frequency near the resonance of $\chi_\text{ee}^{zz}$ is proportional to $k_x^2$, as explained in Appendix~\ref{app_sym_normal}. Near the BIC, and for small angles, the Q factor is thus given by
\begin{equation}
\label{eq_Qsin}
Q\propto\frac{1}{\sin^2\theta}.
\end{equation}
To understand this relation, consider that in a lossless structure, the energy stored in a resonance can only decrease by radiation, so that $\Gamma$ is related to the coupling coefficient, $d$, between the resonance and the radiated wave by (Appendix~\ref{app_rad})
\begin{equation}
\label{eq_ImWd}
\Gamma=\frac{|d|^2}{2}.
\end{equation}
Note that because of reciprocity, $d$ is also the coupling coefficient to the incident wave~\cite{fan2003}. For the resonance of $\chi_\text{ee}^{zz}$, $d\propto\sin\theta$, since $P_z$ only couples to plane waves through their $z$-oriented electric field, which is proportional to $\sin\theta$, leading to $\Gamma\propto\sin^2\theta$, as is typical for symmetry-protected BICs~\cite{hsu2016}. This vanishing linewidth makes the $R_0$ and $T_0$ conditions converge at normal incidence, in agreement with~\eqref{eq_dw_RT}.

Away from its two tips, at $\theta=0^\circ$ and $\theta=90^\circ$ in Fig.~\ref{fig_PM_AmpPhase}a, the Q factor of the sharp feature remains finite and is proportional to the amplitudes of $\chi_\text{mm}^{yy}$ and $\chi_\text{ee}^{zz}$, as discussed in Appendix~\ref{app_sym_peak}. It diverges at those tips, although the divergence at grazing incidence is trivial and does not correspond to a BIC, since the incident wave then propagates along the metasurface.

\subsection{From Symmetry-Protected BIC to Quasi-BIC}
\label{sec_qbic}

We now consider the transition from a symmetry-protected BIC to a quasi-BIC caused by an in-plane symmetry breaking of the particles~\cite{koshelev2018,overvig2020}. Breaking the mirror symmetry $x\rightarrow -x$, while keeping $z\rightarrow -z$, can be modeled by the bianisotropic susceptibilities $\chi_\text{me}^{yz}$ and $\chi_\text{em}^{zy}$~\cite{achouri2023a}, so that~\eqref{eq_const} becomes
\begin{subequations}
\label{eq_biani}
\begin{align}
M_y&=\mu_0\chi_\text{mm}^{yy}H_{y,\text{av}}+\frac{1}{c}\chi_\text{me}^{yz}E_{z,\text{av}},\\
P_z&=\epsilon_0\chi_\text{ee}^{zz}E_{z,\text{av}}+\frac{1}{c}\chi_\text{em}^{zy}H_{y,\text{av}}.
\end{align}
\end{subequations}
Using $E_z=-\eta_0\sin\theta\,H_y$ for a TM-polarized plane wave, where $\eta_0$ is the impedance of free space, the term $k^2\chi_\text{mm}^{yy}+k_x^2\chi_\text{ee}^{zz}$ in~\eqref{eq_seso} becomes
\begin{equation}
\label{eq_biani_Y}
k^2\left[\chi_\text{mm}^{yy}-\sin\theta\left(\chi_\text{me}^{yz}+\chi_\text{em}^{zy}\right)+\sin^2\theta\,\chi_\text{ee}^{zz}\right].
\end{equation}
For a reciprocal metasurface, $\chi_\text{me}^{yz}=-\chi_\text{em}^{zy}$~\cite{achouri2021}, which implies that these bianisotropic susceptibilities do not affect the scattered fields. This shows that the effect of symmetry breaking cannot be modeled in a such simple way.

Instead, we mimic the asymmetry of the structure with an additional resonance of $\chi_\text{mm}^{yy}$ that has the same resonance frequency as $\chi_\text{ee}^{zz}$. This additional resonance plays the role of the bianisotropic susceptibilities in~\eqref{eq_biani}, since it allows the magnetic field of the incident wave to excite, even at normal incidence, a resonance at the same frequency as $P_z$, same as $\chi_\text{em}^{zy}$ would. For a small asymmetry, characterized by a parameter $\alpha$, the coupling between the $P_z$ mode and the incident wave at normal incidence is proportional to $\alpha$. Hence, the amplitude of this additional resonance, which corresponds to the square of this coupling as in~\eqref{eq_ImWd}, is $A_\alpha\propto\alpha^2$.

Figure~\ref{fig_qbic} shows the resulting response for the case when \mbox{$A_\alpha=A_0\alpha^2$ with $A_0=10^{24}$}. The transmittance is essentially unchanged, except near the tip of the sharp feature at $\lambda=600$~nm and $\theta=0^\circ$, where the $R_0$ condition no longer converges to the $T_0$ condition but remains separated from it by a distance that increases with $\alpha$, as illustrated in Figs.~\ref{fig_qbic}c--e. From~\eqref{eq_dw_RT}, this corresponds to a finite linewidth and thus to a quasi-BIC. Since $\Gamma$ is proportional to $A_\alpha$ at normal incidence (Appendix~\ref{app_sym_qbic}), we have that
\begin{equation}
\label{eq_Qalpha}
Q\propto\frac{1}{\alpha^{2}},
\end{equation}
which is the well-known inverse-square law of quasi-BICs~\cite{koshelev2018}. This behavior is confirmed in Fig.~\ref{fig_qbic}f.

\subsection{Avoided Crossings of the Scattering Zeros}
\label{sec_xx}

\begin{figure*}[t] 
	\centering
	\includegraphics[width=\linewidth]{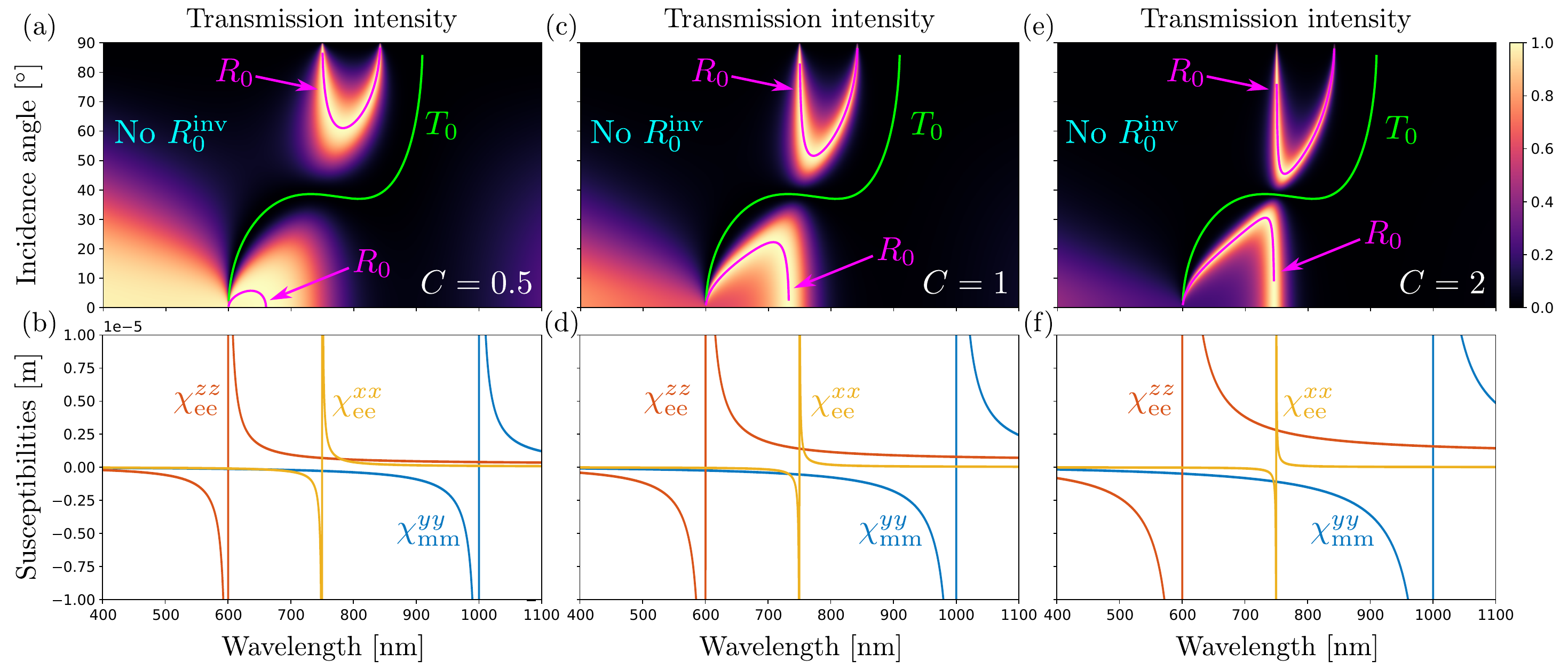}
	\caption{Angular transmittance for a metasurface with the same Lorentzian parameters as those in Fig.~\ref{fig_PM_AmpPhase} but with $\chi_\text{ee}^{xx}$ parametrized by $A_\text{ee}^{xx}=1.5\times 10^{23}$ and $\lambda_\text{ee}^{xx}=750$~nm, and with the susceptibilities renormalized by the constant $C$ as in~\eqref{eq_renorm}, for three values of $C$. The pink and green solid lines denote the $R_{0}$ and $T_{0}$ conditions, respectively. The higher the value of $C$, the higher the Q factor of the response.}
	\label{fig_AC}
\end{figure*}

We now show how the angular scattering response of the metasurface in Fig.~\ref{fig_PM_AmpPhase} can be further controlled by introducing a non-zero $\chi_\text{ee}^{xx}$ susceptibility. One may anticipate that combining the resonances of the three susceptibilities in~\eqref{eq_const} with \emph{positive} amplitudes would produce an accidental BIC. However, we will show that this is not the case and that it rather leads to avoided crossings of the scattering zeros. 

We keep the same parameters for the susceptibilities $\chi_\text{mm}^{yy}$ and $\chi_\text{ee}^{zz}$ as used previously and set $\chi_\text{ee}^{xx}$ to be equal to a Lorentzian resonance at $\lambda_\text{ee}^{xx}=750$~nm, which lies roughly in the middle of the $R_0$ condition in Fig.~\ref{fig_PM_AmpPhase}, with the amplitude given in the caption of Fig.~\ref{fig_AC}. This results in the following consequences: 1) the metasurface cannot exhibit any $R_0^\text{inv}$ condition anymore since $\chi_\text{ee}^{xx} \neq 0$ for all wavelengths, and 2) the divergence of $\chi_\text{ee}^{xx}$ makes the $R_0$ and $T_0$ conditions tend to
\begin{subequations}
\label{eq_S02} 
\begin{align}
    R_{0}:& \quad\sin^2\theta \rightarrow 1\label{eq_R02},\\
    T_{0}:& \quad\sin^2\theta \rightarrow -\frac{\chi_\text{mm}^{yy}}{\chi_\text{ee}^{zz}}.\label{eq_T02}  
\end{align}
\end{subequations}
This means that the divergence of $\chi_\text{ee}^{xx}$ not only breaks apart the $R_0$ condition that was present in Fig.~\ref{fig_PM_AmpPhase}, but also forces a $T_0$ condition right through it since the ratio in~\eqref{eq_T02} leads to a real-valued angle $\theta$. To illustrate this effect, we consider the renormalized susceptibilities
\begin{equation}
\label{eq_renorm}
\chi_\text{mm}^{'yy} = C\chi_\text{mm}^{yy},\quad \chi_\text{ee}^{'xx} = \frac{\chi_\text{ee}^{xx}}{C},\quad \chi_\text{ee}^{'zz} = C\chi_\text{ee}^{zz},
\end{equation}
where $C \in \mathbb{R}^{+}$ is a constant, which allows us to highlight the effect of increasing or decreasing the strength of the susceptibilities. The resulting angular scattering transmittances are plotted for three different values of $C$ in Fig.~\ref{fig_AC}.

The transmittance maps in Fig.~\ref{fig_AC} can be understood from the conditions in~\eqref{eq_S0}, together with the susceptibilities of Fig.~\ref{fig_PM_AmpPhase}c. Sufficiently far from $\lambda=750$~nm, $\chi_\text{ee}^{xx}$ is small compared with $\chi_\text{mm}^{yy}$ and $\chi_\text{ee}^{zz}$, so that the $R_0$ condition in~\eqref{eq_R0} remains close to the $R_0^\text{inv}$ condition of Fig.~\ref{fig_PM_AmpPhase} and follows the same oblique line, going from $\lambda=600$~nm at normal incidence to $\lambda\approx 842$~nm at grazing incidence. Close to $\lambda=750$~nm, in contrast, $\chi_\text{ee}^{xx}$ diverges, and the $R_0$ condition departs from this oblique line. On the short-wavelength side of the resonance, it bends towards normal incidence, which it reaches where $\chi_\text{ee}^{xx}=\chi_\text{mm}^{yy}$, while, on the long-wavelength side, it bends towards grazing incidence, which it reaches at the resonance, according to~\eqref{eq_R02}. The $R_0$ condition is thus split into two branches, one at small angles below $\lambda=750$~nm and one at large angles above it. Between these two branches, i.e., where the oblique line of Fig.~\ref{fig_PM_AmpPhase} would cross the resonance of $\chi_\text{ee}^{xx}$, the $T_0$ condition in~\eqref{eq_T02} passes at the angle given by $\sin^2\theta=-\chi_\text{mm}^{yy}/\chi_\text{ee}^{zz}$ at $\lambda=750$~nm, which corresponds to $\theta\approx 38^\circ$. This leads to the \emph{avoided crossing} visible in Fig.~\ref{fig_AC}, which is due to the fact that the $R_0$ and $T_0$ conditions are mutually exclusive.

As $C$ increases, the resonance of $\chi_\text{ee}^{'xx}$ becomes narrower, so that the two $R_0$ branches get closer to each other, and the $T_0$ condition gets more and more squeezed between them, which increases the Q factor of the overall response, in agreement with~\eqref{eq_dw_RT}. Finally, at the resonance of $\chi_\text{ee}^{zz}$ and $\theta=0^\circ$, we again retrieve the convergence of the $T_0$ and $R_0$ conditions of Fig.~\ref{fig_PM_AmpPhase}, which corresponds to the symmetry-protected BIC.

The Q factor of the squeezed feature can be approximated from the pole of $s_\text{e}$ in~\eqref{eq_seso}. As shown in Appendix~\ref{app_xxQ}, we find that
\begin{equation}
\label{eq_Qxx}
Q\approx\frac{2\omega_0^2C}{A_\text{ee}^{xx}k_z},
\end{equation}
where $\omega_0$ is the resonance frequency of $\chi_\text{ee}^{xx}$. The Q factor increases as $C$ increases or as $A_\text{ee}^{xx}$ decreases. In both cases, this corresponds to reducing the strength of $\chi_\text{ee}^{'xx}$, with respect to the other susceptibilities, and thus to the radiation of the corresponding resonance.

This example shows that combining resonances of $\chi_\text{ee}^{zz}$, $\chi_\text{mm}^{yy}$ and $\chi_\text{ee}^{xx}$ with positive amplitudes leads to avoided crossings of the scattering zeros, which can be used to control the Q factor, but not to a convergence of the $T_0$ and $R_0$ conditions at oblique incidence, i.e., not to an accidental BIC. This is because the linewidth of the resonance of $\chi_\text{ee}^{xx}$, which is proportional to $k_z$, only vanishes at grazing incidence. Forcing an $R_0^\text{inv}$ condition in the middle of the avoided-crossing region, which requires $\chi_\text{ee}^{xx}$ to vanish there, does not lead to such a convergence either, but to a double avoided crossing, as shown in Appendix~\ref{app_DC}.

\subsection{Accidental BIC and Negative Lorentzian Amplitudes}
\label{sec_accidental}

We now look for a configuration where the $R_0$ and $T_0$ conditions converge, in a limiting sense, i.e., without ever crossing, as already discussed in~\secref{sec_zeros}, but this time at an oblique incidence angle. To help identify such convergence conditions, we rewrite~\eqref{eq_S0} as
\begin{subequations}
\label{eq_S03} 
\begin{align}
    R_{0}:& \quad\sin^2\theta =\frac{1}{1+\frac{\chi_\text{ee}^{zz}}{\chi_\text{ee}^{xx}}} - \frac{\chi_\text{mm}^{yy}}{\chi_\text{ee}^{zz}}\left(\frac{1}{\frac{\chi_\text{ee}^{xx}}{\chi_\text{ee}^{zz}}+1}\right)\label{eq_R03},\\
    T_{0}:& \quad\sin^2\theta = -\frac{4}{k^2\chi_\text{ee}^{xx}\chi_\text{ee}^{zz}} -\frac{\chi_\text{mm}^{yy}}{\chi_\text{ee}^{zz}},\label{eq_T03}  
\end{align}
\end{subequations}
where we consider $\chi_\text{ee}^{xx}\neq 0$, which excludes invisibility.

\begin{figure}[t]
\centering
\includegraphics[width=\linewidth]{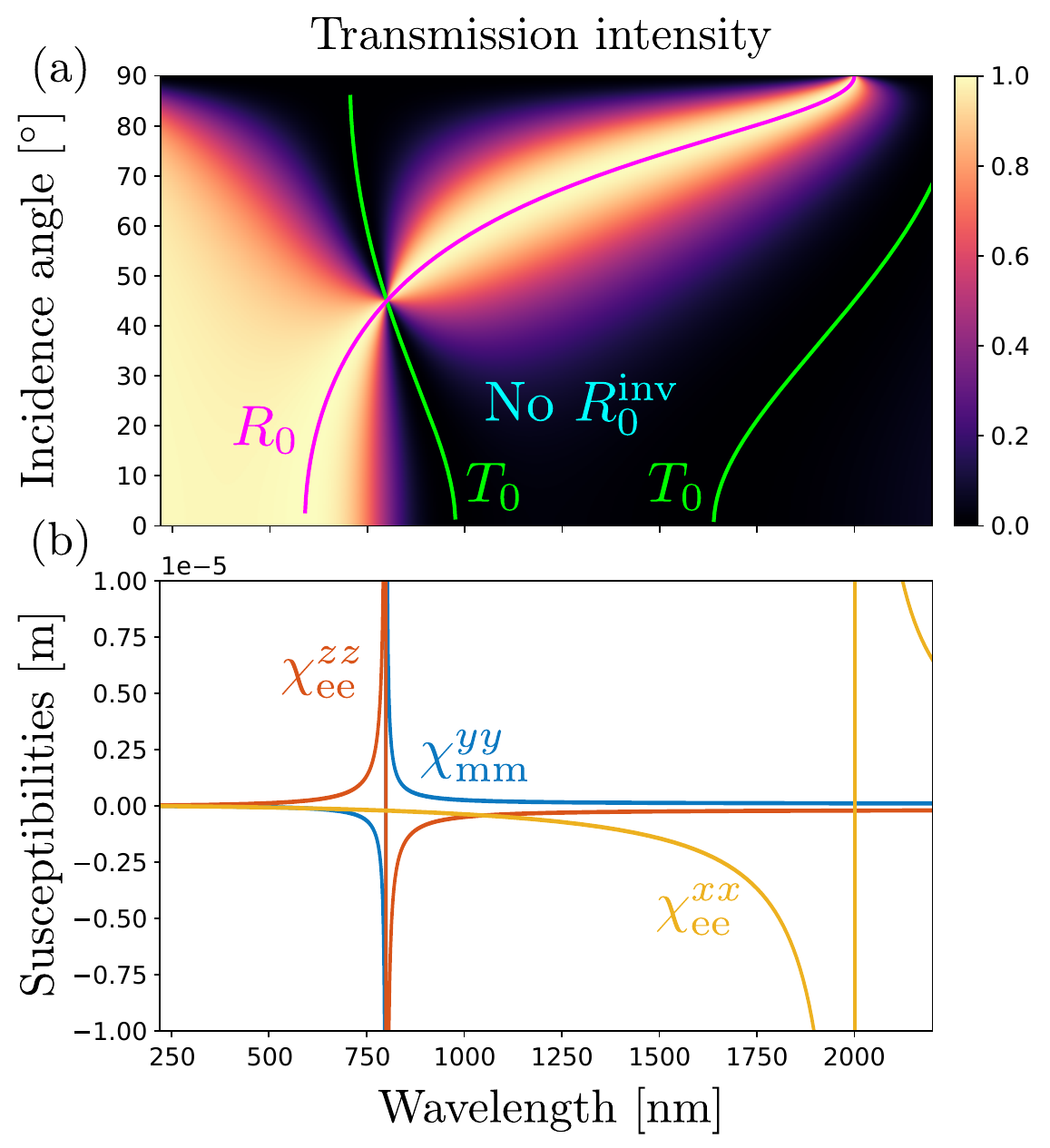}
\caption{ Modeling of an accidental BIC through the convergence of the $T_0$ and $R_0$ conditions. The Lorentzian parameters of the susceptibilities are $A_\text{ee}^{zz}=-10^{24}$, $A_\text{mm}^{yy}=5\times 10^{23}$, $A_\text{ee}^{xx}=10^{24}$, $\lambda_\text{ee}^{zz}=\lambda_\text{mm}^{yy}=800$~nm, and $\lambda_\text{ee}^{xx}=2000$~nm. (a) Transmittance map versus incidence angle and wavelength. (b) Real parts of the susceptibilities.}
\label{fig_C}
\end{figure}

Both expressions contain the common ratio $-\chi_\text{mm}^{yy}/\chi_\text{ee}^{zz}$, so that the two conditions converge if the other terms vanish. This happens when $\chi_\text{ee}^{zz}$ and $\chi_\text{mm}^{yy}$ share the same resonance wavelength, such that both diverge with the same Lorentzian denominator and their ratio tends to the ratio of their amplitudes. The $R_0$ and $T_0$ conditions then converge to
\begin{equation} 
\label{eq_crossing} 
R_0 \quad\text{and}\quad T_0:\quad \sin^{2} \theta \rightarrow -\frac{\chi_\text{mm}^{yy}}{\chi_\text{ee}^{zz}} \rightarrow -\frac{A_\text{mm}^{yy}}{A_\text{ee}^{zz}},
\end{equation}
For this angle to be real, the two amplitudes must have opposite signs. For illustration, we choose a negative amplitude for $\chi_\text{ee}^{zz}$, set the resonances of $\chi_\text{ee}^{zz}$ and $\chi_\text{mm}^{yy}$ at $\lambda=800$~nm and that of $\chi_\text{ee}^{xx}$ at $\lambda=2000$~nm. As can be seen in Fig.~\ref{fig_C}, these conditions lead to a convergence at $\theta=45^\circ$, and produce a high-Q feature at that corresponds to an accidental BIC.

At this convergence point,~\eqref{eq_crossing} breaks down, so that energy conservation is not violated. At the angle given by~\eqref{eq_crossing}, $k^2\chi_\text{mm}^{yy}+k_x^2\chi_\text{ee}^{zz}=0$ for all wavelengths, so that $s_\text{o}=1$ and~\eqref{eq_RTeo} reduces to
\begin{subequations}
\begin{align}
R &= -\frac{k_z\chi_\text{ee}^{xx}}{2j-k_z\chi_\text{ee}^{xx}},\\
T &= \frac{2j}{2j-k_z\chi_\text{ee}^{xx}},
\end{align}
\end{subequations}
whereas approaching this point at the resonance wavelength gives $s_\text{o}=-1$ leading to different scattering parameters, which are thus discontinuous at the accidental BIC~\cite{shipman2012}. 

Two superimposed resonances of the normal electric and the tangential magnetic susceptibility with opposite amplitudes thus correspond to a second mechanism for achieving a BIC, this time an accidental one. However, this is only possible with a negative Lorentzian amplitude, which comes with a consequence: for a lossy ($\gamma \neq 0$) system, a negative amplitude flips the sign of the imaginary part of~\eqref{eq_lorentz},  
\begin{equation} 
\label{eq_lorentzImag} 
    \text{Im}\{\chi_\text{aa}^{bb}(\omega)\} = -\frac{A_\text{aa}^{bb}\omega\gamma}{(\omega_0^2 - \omega^2)^{2} + \omega^{2}\gamma^{2}}.
\end{equation}
which transforms loss into gain. But, even for a lossless ($\gamma = 0$) system, the pole of $s_\text{o}$, obtained in Appendix~\ref{app_acc} for $\chi_\text{ee}^{zz}$ and $\chi_\text{mm}^{yy}$ sharing the resonance frequency $\omega_0$, has an imaginary part
\begin{equation}
\label{eq_ImW}
\Gamma\approx\frac{k^2\left(A_\text{mm}^{yy}+A_\text{ee}^{zz}\sin^2\theta\right)}{4\omega_0k_z}.
\end{equation}
This expression vanishes at the angle given by~\eqref{eq_crossing}, which confirms that it is indeed a BIC and, from~\eqref{eq_dw_RT}, explains the convergence of the zeros. However, this expression also changes sign at the BIC, which implies that the resonance is damped for $\theta<\theta_\text{BIC}$ but grows in time for $\theta>\theta_\text{BIC}$, which corresponds to an active system. Note that this cannot be seen in the transmittance of Fig.~\ref{fig_C}a, since real-valued susceptibilities always lead to $|R|^2+|T|^2=1$. But it clearly corresponds to a violation of Foster's reactance theorem~\cite{foster1924}, according to which the susceptibilities of a lossless and passive local response must increase with frequency and must therefore have positive Lorentzian amplitudes.

Hence, within the proposed dipolar GSTC framework, an accidental BIC can only be achieved with an active metasurface. However, this seems to be in contradiction with the response of the metasurface shown in Fig.~\ref{fig_BIC} that clearly exhibits an accidental BIC even though it is passive and lossless. As we will discuss in the following, it turns out that an accidental BIC is impossible for any passive, lossless and local dipolar metasurface, whatever the number of Lorentzian resonances. For such a metasurface, Foster's theorem implies, as mentioned above, that all the Lorentzian amplitudes are positive. 

Moreover, since the metasurface is lossless, the term $k^2\chi_\text{mm}^{yy}+k_x^2\chi_\text{ee}^{zz}$ in~\eqref{eq_seso} is real for real frequencies, so that the denominator of $s_\text{o}$, $2k_z+j(k^2\chi_\text{mm}^{yy}+k_x^2\chi_\text{ee}^{zz})$, cannot vanish on the real frequency axis for $k_z>0$. In this term, a resonance shared by both $\chi_\text{mm}^{yy}$ and $\chi_\text{ee}^{zz}$ contributes with a strength proportional to $A_\text{mm}^{yy}+A_\text{ee}^{zz}\sin^2\theta$, which sets its linewidth, as in~\eqref{eq_ImW}. A BIC therefore requires this term to vanish, in which case the resonance no longer couples to the incident and radiated waves and disappears from the response. With positive amplitudes, this only happens at normal incidence when $A_\text{mm}^{yy}=0$, i.e., for the symmetry-protected BIC, while the resonance of $s_\text{e}$ only vanishes at grazing incidence, since it is weighted by $k_z=k\cos\theta$. This also holds when several resonances interfere through the same radiation channel.

To understand the physical meaning of the negative amplitude, consider that the two superimposed resonances correspond to a single mode with both $M_y$ and $P_z$ responses. These polarizations belong to the odd part of the response and therefore radiate into the same channel, unlike $P_x$. However, since they behave differently under $x\rightarrow -x$, they can only be combined at oblique incidence. It follows that the accidental BIC corresponds to the angle at which their radiated fields cancel by destructive interference. In the susceptibility model, a single mode corresponds to Lorentzian resonances that share the same pole and whose amplitudes are products of the coupling coefficients of the mode, in contrast with independent resonances, whose amplitudes simply add.

In a passive structure, the radiation rate, which is here equivalent to the radiation loss of this mode, is given by~\eqref{eq_ImWd} and can never be negative but must still vanish at the BIC. It must therefore reach a minimum at $\theta_\text{BIC}$. This means that the radiation rate should be, in first approximation, of the form $\Gamma\propto(\theta-\theta_\text{BIC})^2$, which leads to $Q\propto(\theta-\theta_\text{BIC})^{-2}$~\cite{jin2019,yuan2018}, as confirmed in~\subsecref{sec_ridge} for the metasurface of Fig.~\ref{fig_BIC}. It turns out, however, that in our dipolar model, the contributions of $M_y$ and $P_z$ to $\Gamma$ simply add, since these polarizations are excited independently by $H_y$ and $E_z$ in~\eqref{eq_const}. This means that they can only cancel if $A_\text{mm}^{yy}$ and $A_\text{ee}^{zz}$ have opposite signs, which is how the dipolar model represents the destructive interference between their radiated fields. Moreover, since $A_\text{mm}^{yy}+A_\text{ee}^{zz}\sin^2\theta$ is linear in $\sin^2\theta$, it changes sign at $\theta_\text{BIC}$ instead of reaching a minimum, which is why the model becomes active above the BIC. This also implies that the Q factor has a distinct dependence on the angle, $Q\propto|\theta-\theta_\text{BIC}|^{-1}$, as explained in Appendix~\ref{app_acc}. 

A passive accidental BIC instead requires a radiation rate proportional to $(\sin^2\theta-\sin^2\theta_\text{BIC})^2$, which contains a term in $\sin^4\theta$, i.e., in $k_x^4$. Such a term does not exist in the dipolar GSTC model, where the strength of the resonance only depends on the angle through $k_x^2\chi_\text{ee}^{zz}$~\cite{achouri2020a}, but it may, for instance, be provided by higher-order multipolar responses~\cite{achouri2022b,allahverdizadeh2025a}. As shown in Appendix~\ref{app_multipoles}, the electric quadrupole $Q_{xz}^{\mathrm{(e)}}$ and the magnetic octupole $O_{yzz}^{\mathrm{(m)}}$ can couple to the incident wave with the factors $\cos(2\theta)$ and $\cos^2\theta$, respectively, and thus contribute to the angular dependence of the coupling coefficient, which a dipolar model can only attribute to $\chi_\text{ee}^{zz}$.

This suggests that if the susceptibilities of the dipolar model were interpreted as effective susceptibilities that incorporate higher-order multipolar contributions, the negative amplitude of the Lorentzian function could simply correspond to the combined effect of these responses~\cite{achouri2022b,dugan2022a,tiukuvaara2023}. Note that similar higher-order terms in $k_x$ can also originate from the lattice coupling between the particles of the array, even if each particle is purely dipolar, which explains why accidental BICs can be obtained with coupled-dipole models~\cite{abujetas2020,abujetas2022}. In both cases, the effective dipolar susceptibilities may exhibit negative amplitudes even though the metasurface is passive and lossless, similarly to the non-physical effective parameters known in the retrieval of metamaterial parameters~\cite{koschny2003,Alu2011}. However, as we will show in details in Sec.~\ref{sec_passive} using a simple passive nonlocal model, accidental BICs in metasurfaces are intrinsically nonlocal because they require terms in $k_x^4$. In contrast, the symmetry-protected BICs can be local, since $\Gamma\propto\sin^2\theta$ already reaches a minimum at normal incidence.

\begin{figure}[h!]
\centering
\includegraphics[width=0.9\linewidth]{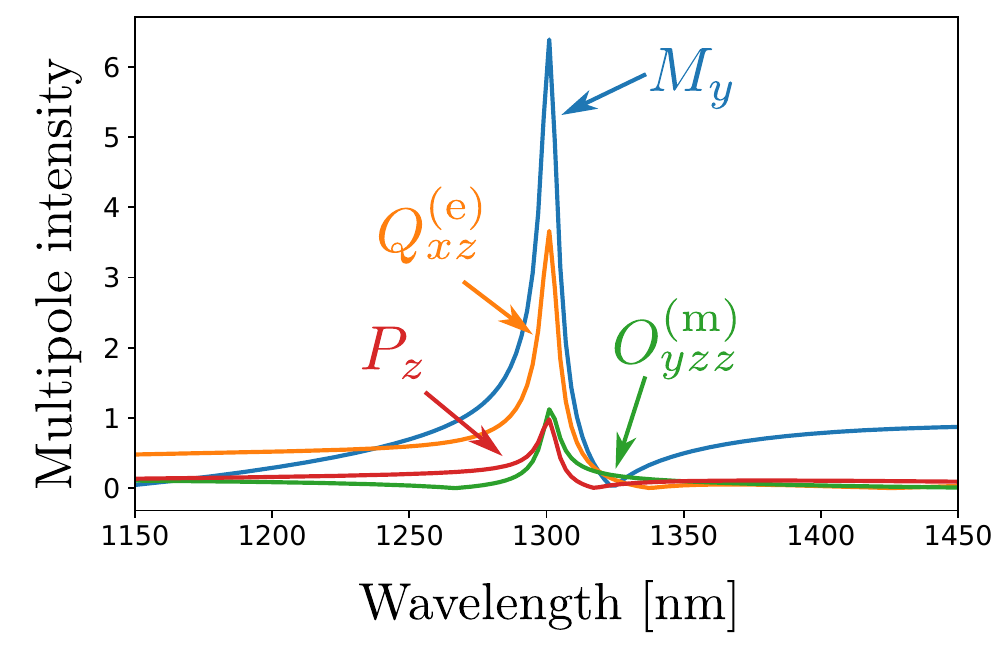}
\caption{Scattering contributions of the dominant multipole moments of the silicon-ridge metasurface of Fig.~\ref{fig_BIC} for an incidence angle of $\theta=15^\circ$, computed using the method of~\cite{allahverdizadeh2025a}.}
\label{fig_multipoles}
\end{figure}

To verify that the response of the metasurface of Fig.~\ref{fig_BIC} involves higher-order multipoles, we perform a multipolar decomposition at $\theta=15^\circ$, slightly below the BIC angle, shown in Fig.~\ref{fig_multipoles}. As can be seen, the dominant contributions are the magnetic dipole $M_y$, the electric dipole $P_z$, the electric quadrupole $Q_{xz}^{\mathrm{(e)}}$ and the magnetic octupole $O_{yzz}^{\mathrm{(m)}}$, which all change sign under $z\rightarrow -z$ and thus belong to the odd part of the response, which is consistent with the dipolar model, where the accidental BIC is a resonance of $s_\text{o}$. The quadrupole and octupole are comparable to the dipoles, so that the response near the BIC cannot be described by dipoles alone, while $P_z$ is weak, which makes sense since it only appears at oblique incidence. The angular dependence of the coupling coefficient is thus expected to be mainly provided by the quadrupole and the octupole, which, as discussed above, a dipolar model can only attribute to $\chi_\text{ee}^{zz}$. As shown in Sec.~\ref{sec_passive}, the effective dipolar description of this metasurface indeed leads to a large negative amplitude of $\chi_\text{ee}^{zz}$. Note that these multipolar contributions originate from the retardation effect, which is significant since the height of                            the ridges is comparable with the wavelength. Nevertheless, the metasurface can be modeled as a zero-thickness sheet, since its period is subwavelength, provided that these multipolar, or equivalently spatially dispersive, contributions are included in the GSTC~\cite{achouri2022b,tiukuvaara2023,allahverdizadeh2025a}, as we will show in Sec.~\ref{sec_passive}.

\section{A Passive Spatially Dispersive Model of Accidental BICs}
\label{sec_passive}

\begin{figure*}[t!]
	\centering
	\includegraphics[width=\linewidth]{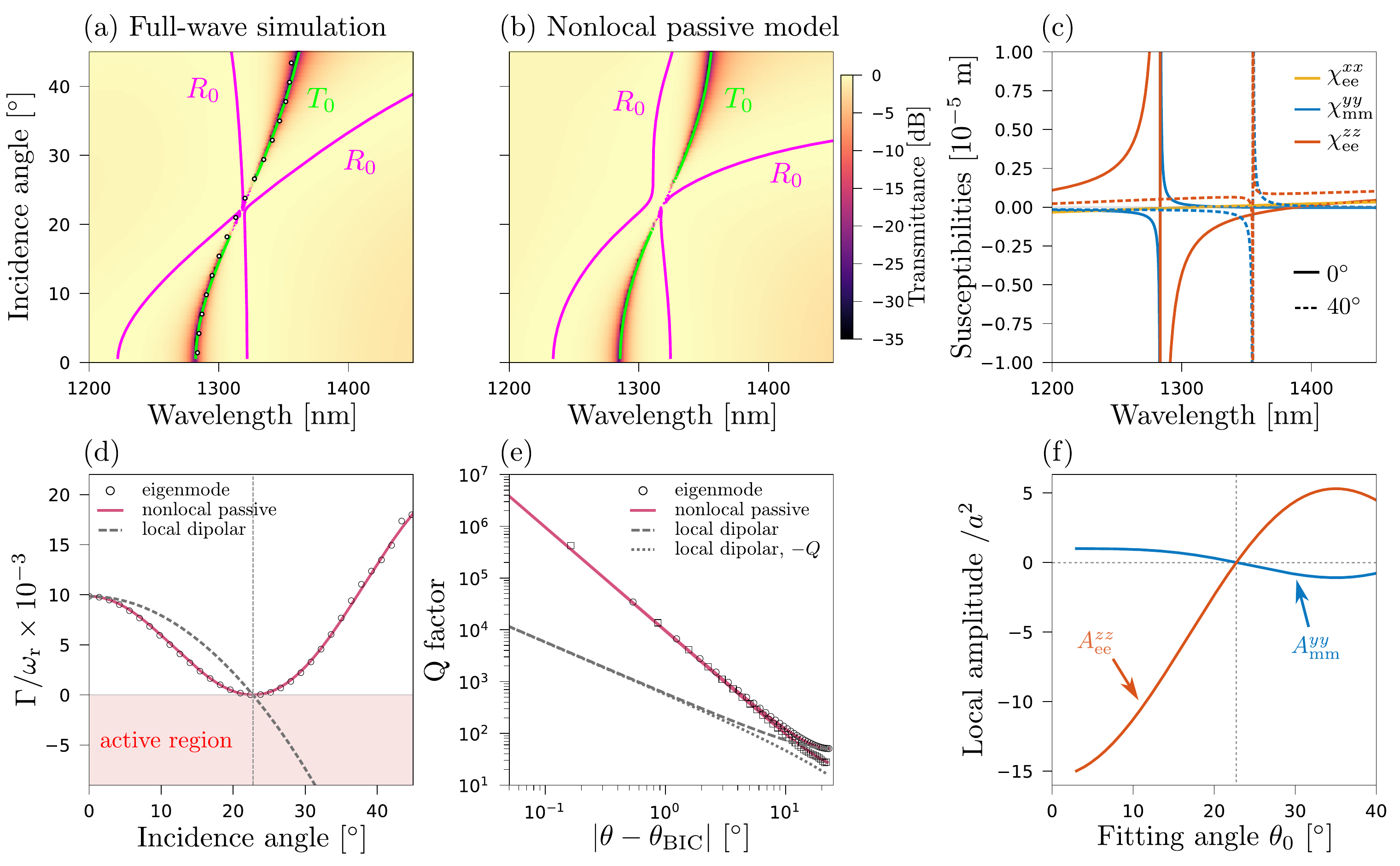}
	\caption{Passive spatially dispersive (nonlocal) model of the accidental BIC of the silicon-ridge metasurface of Fig.~\ref{fig_BIC}, with parameters extracted from the eigenmode analysis of~\subsecref{sec_ridge}. (a) Full-wave transmittance, with the eigen-wavelengths of the resonant mode (white markers), and (b) transmittance of the model. The pink and green solid lines denote the $R_0$ and $T_0$ conditions, respectively. (c) Real parts of the susceptibilities of the model at $\theta=0^\circ$ (solid) and $\theta=40^\circ$ (dashed). (d) Imaginary part of the resonance frequency, normalized by its real part, of the eigenmode (circles), of the nonlocal passive model (solid), and of a local dipolar model having its BIC at the same angle (dashed). (e) Q factor versus the distance to the BIC, $|\theta-\theta_\text{BIC}|$, of the eigenmode below (circles) and above (squares) the BIC, of the nonlocal passive model (solid), and of the dipolar model (dashed), whose Q factor is negative above the BIC (dotted). Each Q factor is plotted with respect to its own BIC angle. (f) Amplitudes of the dipolar model fitted locally, at the angle $\theta_0$, to the strength of the passive model, normalized by $a^2$. The dotted lines indicate $\theta_\text{BIC}=22.7^\circ$. The extraction of the parameters of the model is shown in Fig.~\ref{fig_fit}.}
	\label{fig_passive}
\end{figure*}

In~\subsecref{sec_accidental}, we have seen that a passive accidental BIC requires a term in $\sin^4\theta$, i.e., in $k_x^4$, which may be provided by multipolar contributions or lattice coupling. We now show that a simple spatially dispersive extension of the dipolar model indeed leads to a passive accidental BIC.

\subsection{Minimal Passive Model}
\label{sec_minimal}

We consider the same configuration as in Fig.~\ref{fig_C}, but with a normal susceptibility that depends on the tangential wavevector,
\begin{subequations}
\label{eq_passive_chi}
\begin{align}
\chi_\text{mm}^{yy}&=\frac{A_\text{mm}^{yy}}{\Omega^2-\omega^2},\\
\chi_\text{ee}^{zz}&=\frac{A_\text{ee}^{zz}+A_\text{ee}^{zz,(2)}\,k_x^2/k^2}{\Omega^2-\omega^2},
\end{align}
\end{subequations}
where $A_\text{ee}^{zz,(2)}$ is the amplitude of the spatially dispersive part of $\chi_\text{ee}^{zz}$, and where the resonance frequency, $\Omega^2=\omega_0^2-\eta c^2k_x^2$ with $\eta$ a real constant, accounts for the band dispersion of the mode, which red-shifts with the angle for $\eta>0$. Such an angle-dependent resonance also occurs in a simple array of purely electric-dipolar metallic rods~\cite{rahmeier2022a}, and it provides a better model of the dispersive nature of the mode in Fig.~\ref{fig_BIC}. The term $k^2\chi_\text{mm}^{yy}+k_x^2\chi_\text{ee}^{zz}$ in~\eqref{eq_seso} then becomes $k^2W(\theta)/(\Omega^2-\omega^2)$, where the strength of the resonance, $W(\theta)$, combines the Lorentzian amplitudes of $\chi_\text{mm}^{yy}$ and $\chi_\text{ee}^{zz}$ as
\begin{equation}
\label{eq_Atheta}
W(\theta)=A_\text{mm}^{yy}+A_\text{ee}^{zz}\sin^2\theta+A_\text{ee}^{zz,(2)}\sin^4\theta,
\end{equation}
which is proportional to $\Gamma$, as shown in Appendix~\ref{app_passiveQ}. For a passive metasurface, $W(\theta)$ can never be negative, and must therefore vanish at an accidental BIC without changing sign. This means that it must be at least a quadratic function. Since~\eqref{eq_Atheta} is quadratic in $\sin^2\theta$, this requires its discriminant to vanish, $(A_\text{ee}^{zz})^2=4A_\text{mm}^{yy}A_\text{ee}^{zz,(2)}$, which is easily satisfied by writing $A_\text{mm}^{yy}=a^2$, $A_\text{ee}^{zz}=2ab$ and $A_\text{ee}^{zz,(2)}=b^2$, with $a$ and $b$ real, so that
\begin{equation}
\label{eq_square}
W(\theta)=\left(a+b\sin^2\theta\right)^2,
\end{equation}
which vanishes at $\sin^2\theta_\text{BIC}=-a/b$. Here, $a$ and $b$ have opposite signs, so that $A_\text{ee}^{zz}=2ab$ is negative. With this approach, the negative amplitude of $\chi_\text{ee}^{zz}$ is still present but it is compensated by the positive term $b^2\sin^4\theta$, which is exactly the term proportional to $k_x^4$ that is missing from the dipolar model. Physically,~\eqref{eq_square} corresponds to a single resonant mode, whose radiation rate is the square of its coupling coefficient, as in~\eqref{eq_ImWd}, with $d\propto a+b\sin^2\theta$, where $a$ and $b$ combine the contributions of $M_y$, $P_z$, $Q_{xz}^{\mathrm{(e)}}$ and $O_{yzz}^{\mathrm{(m)}}$, as explained in Appendix~\ref{app_multipoles}. 

Note that the model in~\eqref{eq_passive_chi} is only a first-order approximation. More generally, one may express the strength of a single resonant mode as $W(\theta)=d^2(\theta)$, where $d$ is a real function of $\sin^2\theta$. The strength of the resonance is then never negative, so that the model is passive for any $d$, and an accidental BIC occurs whenever $d=0$.

\subsection{Application to the Silicon-Ridge Metasurface}
\label{sec_ridge}

\begin{figure*}[t!]
	\centering
	\includegraphics[width=\linewidth]{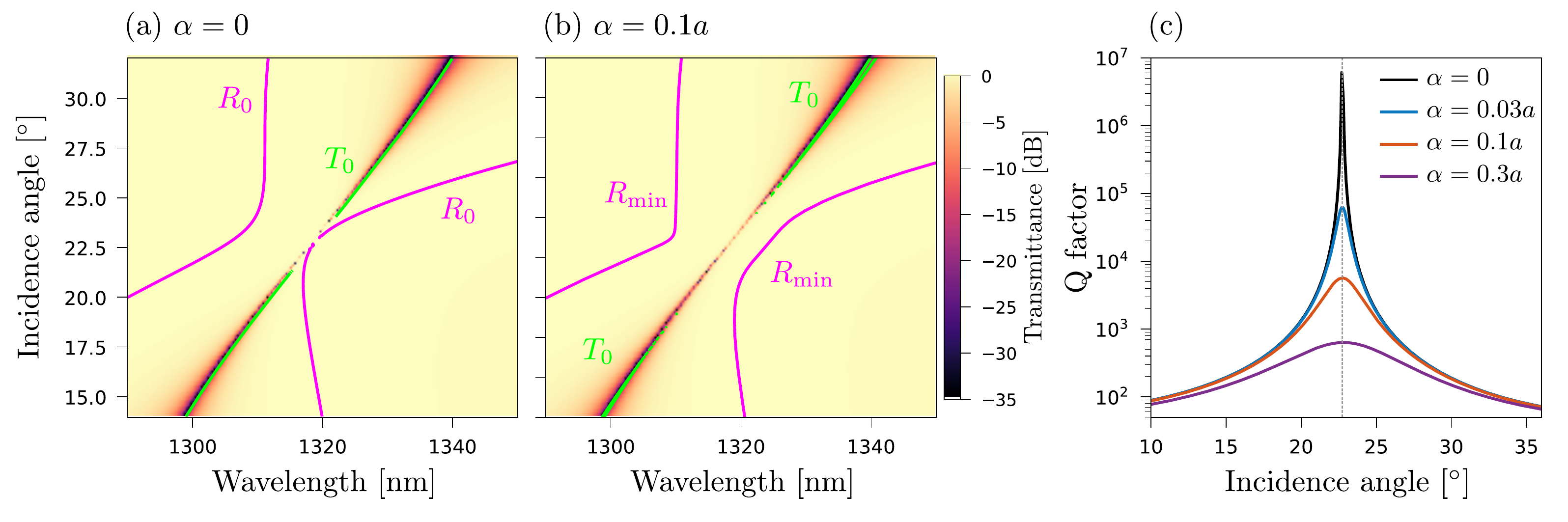}
	\caption{Effect of breaking the mirror symmetry $z\rightarrow -z$ on the accidental BIC of the silicon-ridge model of~\subsecref{sec_ridge}, described by the omega-type bianisotropic susceptibilities in~\eqref{eq_zbreak}. Transmittance (a) without asymmetry ($\alpha=0$) and (b) with $\alpha=0.1a$. The pink and green solid lines denote the $R_0$ and $T_0$ conditions in (a), while, in (b), they denote the loci of the reflection minima and the $T_0$ conditions, respectively, since exact reflection zeros then only exist at isolated points. (c) Q factor of the resonance versus the angle of incidence for different values of $\alpha$. The dotted line indicates $\theta_\text{BIC}=22.7^\circ$.}
	\label{fig_zbreak}
\end{figure*}

We now apply this modeling approach to the silicon-ridge metasurface of Fig.~\ref{fig_BIC}. For this purpose, we first perform an eigenmode analysis of this metasurface with COMSOL, which provides the complex frequency of its quasi-normal mode versus the angle. We then overlay the eigen-wavelengths of its resonant mode on the transmittance map in Fig.~\ref{fig_passive}a, which is a zoomed-in version of Fig.~\ref{fig_BIC}c. We also show that the incident wave strongly couples to this mode at normal incidence but no longer couples to it at the BIC, as expected for a BIC and as illustrated in Appendix~\ref{app_fields}. Moreover, close to the BIC, the Q factor of the eigenmode scales as $Q\propto(\theta-\theta_\text{BIC})^{-2}$, as can be seen in Fig.~\ref{fig_passive}e, so that the radiation rate of the physical structure vanishes quadratically, as expected for a passive structure, in contrast with the dipolar model of~\subsecref{sec_accidental}, whose radiation rate vanishes linearly and changes sign.

Next, we extract the parameters of the resonance dispersion directly from the eigenfrequencies, $\tilde{\omega}=\omega_\text{r}+j\Gamma$, by fitting them, as detailed in Appendix~\ref{app_fit}. However, it turns out that the first-order model in~\eqref{eq_passive_chi} is not sufficient to properly capture the response of the structure in Fig.~\ref{fig_BIC}. For this reason, we consider the next order in $\sin^2\theta$, both in the angular dispersion and in the coupling coefficient. The angular dispersion of the resonance is then well described by $\Omega^2=\omega_0^2-(\eta\sin^2\theta+\zeta\sin^4\theta)\omega^2$, where $\eta$ and $\zeta$ are real parameters and where we have found that $\lambda_0=1283.4$~nm, $\eta=0.429$ and $\zeta=-0.367$. Similarly, the coupling coefficient that we extracted from the linewidth of the eigenmode is well described by $d=(a+b\sin^2\theta)\cos^2\theta$, which corresponds to~\eqref{eq_square} multiplied by $\cos^2\theta=1-\sin^2\theta$. As explained in Appendix~\ref{app_multipoles}, this is a higher-order term that may result from the angular variation of the multipole moments of the mode. Note that it preserves passivity since $W(\theta)=d^2$ remains satisfied. Using the fitted data, we find that the angle at which the BIC occurs is \mbox{$\theta_\text{BIC}=22.7^\circ$}, which is close to the actual value of $22.9^\circ$ obtained from the eigenmode analysis.

The model described above provides only the resonant parts of $\chi_\text{mm}^{yy}$ and $\chi_\text{ee}^{zz}$. To complete it, we add non-resonant background contributions to $\chi_\text{ee}^{xx}$, $\chi_\text{mm}^{yy}$ and $\chi_\text{ee}^{zz}$, as detailed in Appendix~\ref{app_fit}. The resulting susceptibilities are plotted in Fig.~\ref{fig_passive}c. We see that $\chi_\text{ee}^{xx}$ is non-resonant in this wavelength range, while the resonances of $\chi_\text{mm}^{yy}$ and $\chi_\text{ee}^{zz}$ red-shift with the angle. The corresponding transmittance of the model is plotted in Fig.~\ref{fig_passive}b, and reproduces very well the full-wave one shown in Fig.~\ref{fig_passive}a. The model also reproduces rather well the behavior of the scattering zeros. Indeed, the $T_0$ condition follows the transmission dip between two $R_0$ conditions, and the $R_0$ condition on its long-wavelength side converges onto it at the BIC, which corresponds to the convergence of scattering zeros discussed in~\secref{sec_zeros} but for a passive structure. Note that in the full-wave response, the second $R_0$ condition also passes through the BIC, whereas in the model it remains more than $10$~nm away from it, which is due to the approximate description of the non-resonant background contributions.

Figures~\ref{fig_passive}d and~\ref{fig_passive}e compare $\Gamma$ and the Q factor of this passive nonlocal model with those of the eigenmode and of a local dipolar model with the same BIC angle, i.e., with a strength proportional to $1-\sin^2\theta/\sin^2\theta_\text{BIC}$. The passive nonlocal model follows the eigenmode over the whole angular range, with a value of $\Gamma$ that only touches zero at the BIC where we have $Q\propto(\theta-\theta_\text{BIC})^{-2}$. In the local dipolar model, $\Gamma$ changes sign at the BIC, so that $Q\propto|\theta-\theta_\text{BIC}|^{-1}$ and becomes negative above the BIC, which indicates growing fields.

Finally, we use this model to explain why the effective dipolar susceptibilities of a passive accidental BIC exhibit amplitudes of opposite signs. As explained previously, the local dipolar model describes the strength of the resonance as $A_\text{mm}^{yy}+A_\text{ee}^{zz}\sin^2\theta$, which is a linear function of $\sin^2\theta$, whereas the strength of the nonlocal passive model, $W(\theta)$, is not. We now try to connect the two models together by adjusting the amplitudes $A_\text{ee}^{zz}$ and $A_\text{mm}^{yy}$ of the local dipolar model so that they lead to a response that is similar to that of the nonlocal passive model. Naturally, this means that these amplitudes must become angle-dependent so that they can be matched at every angle. To do so, we consider an angle $\theta_0$ around which the function $W(\theta)$ is expanded in a Taylor series in terms of $\sin^2\theta$. We use the substitutions $u=\sin^2\theta$ and $u_0=\sin^2\theta_0$ to obtain $W(u)\approx W(u_0)+(u-u_0)\,\mathrm{d}W/\mathrm{d}u|_{u_0}$, which has the same form as the local dipolar model $A_\text{mm}^{yy}+A_\text{ee}^{zz}u$ but with the amplitudes given by
\begin{subequations}
\label{eq_local_amp}
\begin{align}
A_\text{ee}^{zz}(\theta_0)&=\left.\frac{\mathrm{d}W}{\mathrm{d}u}\right|_{u_0},\\
A_\text{mm}^{yy}(\theta_0)&=W(u_0)-u_0\left.\frac{\mathrm{d}W}{\mathrm{d}u}\right|_{u_0}.
\end{align}
\end{subequations}
These amplitudes are plotted in Fig.~\ref{fig_passive}f. Since $W(\theta)$ is positive and reaches its minimum, zero, at the BIC, its slope is negative below the BIC and positive above it. Below the BIC, $A_\text{ee}^{zz}$ is therefore negative, as in Fig.~\ref{fig_C}, while $A_\text{mm}^{yy}$ remains positive. Above the BIC, it is the opposite, $A_\text{ee}^{zz}$ becomes positive, while $A_\text{mm}^{yy}$ becomes negative. In both cases, the two amplitudes have opposite signs at every angle. This mapping clearly demonstrates that the negative amplitude of the local dipolar model is a linear approximation of a passive and spatially dispersive response.

\subsection{Breaking the Mirror Symmetry $z\rightarrow -z$}
\label{sec_zbreak}

In our description of the accidental BIC, the silicon-ridge metasurface satisfied the mirror symmetry $z\rightarrow -z$, which ensures that the resonant mode only radiates into a single channel. We now investigate the effect of breaking this symmetry on the existence of the accidental BIC, which can be achieved by giving the ridges a trapezoidal cross-section or, for three-dimensional particles, by transforming cylinders into truncated cones (similar to the earlier work~\cite{mikheeva2023} in the context of behavior of the poles and zeros). Naturally, the presence of a substrate also breaks this symmetry, but it additionally makes the background media different on both sides of the metasurface, which leads to a more involved GSTC model that differs from the one in~\eqref{eq_RT}. To keep the same modeling approach, we thus consider a particle with broken symmetry rather than a substrate. In the GSTC, such a structural asymmetry is described by the omega-type bianisotropic susceptibilities $\chi_\text{em}^{xy}=-\chi_\text{me}^{yx}$~\cite{achouri2021,achouri2023a}, which couple the even polarization $P_x$ to the odd polarization $M_y$, and which are purely imaginary for a reciprocal and lossless metasurface.

The particle asymmetry is thus modeled as an additional resonance in the form of a bianisotropic contribution whose amplitude, $A_\text{em}^{xy}$, is proportional to a small parameter $\alpha$ that characterizes the asymmetry. Passivity, however, requires the resonance to have a positive amplitude, as imposed by Foster's theorem, for any combination of the exciting fields $E_x$ and $H_y$. As shown in Appendix~\ref{app_omega_pass}, this leads to the condition
\begin{equation}
\label{eq_passive_cond}
A_\text{ee}^{xx}W(\theta)\geq|A_\text{em}^{xy}|^2,
\end{equation}
where $A_\text{ee}^{xx}$ is the amplitude of a resonance of $\chi_\text{ee}^{xx}$ at the same frequency and $W(\theta)=d^2(\theta)$ is the strength of the resonance of~\subsecref{sec_minimal}, with $d(\theta)=(a+b\sin^2\theta)\cos^2\theta$ for the silicon-ridge metasurface. Somewhat counterintuitively, the only way to add a resonant bianisotropic susceptibility in a passive and lossless fashion is to also consider a resonance of $\chi_\text{ee}^{xx}$, while the bianisotropic amplitude must vanish at the BIC angle, where $d$ vanishes. Thus, the simplest way to satisfy~\eqref{eq_passive_cond} is to consider the case of equality, $A_\text{ee}^{xx}W(\theta)=|A_\text{em}^{xy}|^2$, such that the resonance still corresponds to a single mode, which leads to
\begin{subequations}
\label{eq_zbreak}
\begin{align}
\chi_\text{ee}^{xx}&\rightarrow\chi_\text{ee}^{xx}+\frac{\alpha^2}{\Omega^2-\omega^2},\\
\chi_\text{em}^{xy}&=-\chi_\text{me}^{yx}=\frac{j\alpha\,d(\theta)}{\Omega^2-\omega^2},
\end{align}
\end{subequations}
while the other susceptibilities are unchanged. The resonance of $\chi_\text{ee}^{xx}$ allows the mode to radiate into the even part of the response, and thus to remain coupled to the incident wave at the BIC angle. Because of $\chi_\text{em}^{xy}$, the even and odd parts of the response, $s_\text{e}$ and $s_\text{o}$, are now coupled to each other, and the scattering parameters, derived in Appendix~\ref{app_omega_RT}, remain lossless and reciprocal.

Figure~\ref{fig_zbreak} compares the transmittance of the silicon-ridge model without and with this asymmetry. For $\alpha=0$, the transmission dip disappears at the BIC, whereas, for $\alpha\neq 0$, it remains visible at all angles, since the mode now also radiates into the even part of the response, even when its odd coupling coefficient vanishes. A BIC would then require both its odd and even coupling coefficients, $d$ and the one proportional to $\alpha$, to vanish at the same angle, which is not possible for $\alpha\neq 0$. The accidental BIC thus turns into a quasi-BIC, whose maximum Q factor, shown in Fig.~\ref{fig_zbreak}c, follows the inverse-square law $Q_\text{max}\propto 1/\alpha^2$. The scattering zeros are also strongly affected by the presence of the bianisotropic susceptibilities, as shown in Appendix~\ref{app_omega_zeros}. Indeed, the reflection zeros, which can be deduced from~\eqref{eq_RT_omega}, only occur at isolated points of the wavelength--angle plane. This means that the $R_0$ conditions become reflection \emph{minima}, while the $T_0$ conditions still follow linear dispersion. As shown in Fig.~\ref{fig_zbreak}b, these reflection minima remain separated from the $T_0$ condition around the BIC angle, which is consistent with the finite linewidth of the resonance and~\eqref{eq_dw_RT}. This is thus the counterpart, for the accidental BIC, of the in-plane asymmetry that turns the symmetry-protected BIC into a quasi-BIC in~\subsecref{sec_qbic}.

\section{Conclusion}
\label{sec_conclusion}

We have described BICs in metasurfaces through the interactions of the reflection and transmission zeros of an effective-susceptibility model based on the GSTC. These zeros can never coincide because of energy conservation, but a high-Q response can still be obtained when a transmission zero is squeezed between two reflection zeros, or vice versa. A BIC is then achieved when these zeros converge, which corresponds to a pole and a zero of the scattering matrix merging on the real frequency axis.

We have shown that a symmetry-protected BIC arises at normal incidence from a susceptibility resonance that cannot be excited due to symmetry mismatch, leading to $Q\propto1/\sin^2\theta$, and that an in-plane asymmetry $\alpha$ turns it into a quasi-BIC with $Q\propto1/\alpha^2$. In contrast, combining susceptibility resonances with positive amplitudes only leads to avoided crossings of the zeros, which can be used to control the Q factor but cannot produce an accidental BIC. Within a local dipolar model, an accidental BIC requires two superimposed resonances with amplitudes of opposite signs, which makes the model active on one side of the BIC, and we have shown that this is a general limitation of passive and local dipolar models. Physically, the two resonances represent a single mode carrying both a tangential magnetic and a normal electric response, whose radiated fields cancel at the BIC, and the negative amplitude is the only way such a local dipolar model can represent this interference. To obtain an accidental BIC in a passive and lossless fashion, it is instead required that the strength of the resonance be a function proportional to $\sin^4\theta$. This may be achieved by considering spatially dispersive susceptibilities, either provided by higher-order multipoles or lattice coupling, which leads to $Q\propto(\theta-\theta_\text{BIC})^{-2}$ and suggests that accidental BICs are intrinsically nonlocal.

To verify our analysis, we simulated a lossless metasurface made of silicon ridges that exhibits an accidental BIC. The response of the metasurface near the BIC is strongly multipolar and the corresponding eigenmodes behave as $Q\propto(\theta-\theta_\text{BIC})^{-2}$. To accurately model its Q factor and transmittance, we developed a simple passive and spatially dispersive model based on an extended GSTC formalism, whose parameters were extracted from the full-wave simulations. By mapping the nonlocal susceptibilities of this model to those of the initial local dipolar model, we were able to show that the negative Lorentzian amplitude of one of the local susceptibilities, which violates Foster's reactance theorem, should not be interpreted as gain but as evidence of a nonlocal response~\cite{Alu2011a}. This confirms that accidental BICs should be modeled with spatially dispersive susceptibilities, whose resonance strength must be proportional to $\sin^4\theta$ to satisfy passivity.

Most of our discussion was centered around lossless metasurfaces that are symmetric with respect to $z\rightarrow -z$. This symmetry was considered in the first investigated cases because it ensures that the odd part of the response radiates into a single channel with a real coupling coefficient, so that the condition $d=0$ is sufficient to obtain an accidental BIC~\cite{zhen2014}. However, breaking this symmetry, for instance with tapered resonators, introduces an omega-type bianisotropy that couples together the even and odd parts of the response. This turns the accidental BIC into a quasi-BIC~\cite{yin2020}, whose maximum Q factor follows an inverse-square law related to this asymmetry. Note that such an asymmetry can only be modeled in a passive and lossless fashion provided that a resonance of $\chi_\text{ee}^{xx}$ is included.

Introducing absorption in our model would turn the $R_0$ and $T_0$ conditions into reflection and transmission minima and, close to the BIC, where the radiation rate becomes smaller than the absorption rate, the transmission dip would vanish while the Q factor would saturate at the value set by absorption, since $1/Q=1/Q_\text{rad}+1/Q_\text{abs}$. Loss also makes the limitation of local dipolar models more apparent, since a negative amplitude then corresponds to gain. Finally, extending this framework to more general bianisotropic metasurfaces and to substrated ones, whose background media differ on both sides~\cite{tiukuvaara2023,allahverdizadeh2025a}, would provide further degrees of freedom to control both symmetry-protected and accidental BICs.

\begin{acknowledgments}
We acknowledge funding from the Swiss National Science Foundation (project TMSGI2\_218392).
\end{acknowledgments}

\clearpage
\appendix

\section{Eigenvalues of the Scattering Matrix}
\label{app_S}

In this appendix, we give the physical meaning of the coefficients $s_\text{e}$ and $s_\text{o}$ introduced in Sec.~\ref{sec_model}. Since the metasurface is symmetric with respect to the plane $z=0$, its reflection and transmission coefficients are the same for waves incident from either side. Its scattering matrix, which relates the amplitudes of the waves incident from both sides to the amplitudes of the outgoing waves, thus satisfies the eigenvalue equation
\begin{equation}
\label{eq_Smatrix}
\begin{pmatrix} R & T\\ T & R\end{pmatrix}\begin{pmatrix}1\\ \pm 1\end{pmatrix} = (R\pm T)\begin{pmatrix}1\\ \pm 1\end{pmatrix},
\end{equation}
where the wave amplitudes are defined by their magnetic field $H_y$ at the metasurface, as in~\eqref{eq_RT}. The two eigenvalues of the scattering matrix are therefore $R+T=s_\text{o}$ and $R-T=-s_\text{e}$.

The eigenvector $(1,1)$ corresponds to two waves incident from both sides with the same magnetic field at the metasurface. In that case, $E_{x,\text{av}}=0$ and only $M_y$ and $P_z$ are excited. The eigenvector $(1,-1)$ corresponds to two waves with opposite magnetic fields at the metasurface. In that case, $H_{y,\text{av}}=E_{z,\text{av}}=0$ and only $P_x$ is excited. Finally, since the scattering matrix of a lossless metasurface is unitary, its eigenvalues have unit magnitude, which leads to $|s_\text{e}|=|s_\text{o}|=1$.

\section{Scattering Zeros Near a Resonance}
\label{app_zeros}

In this appendix, we derive~\eqref{eq_dw_RT} of Sec.~\ref{sec_zeros}, which relates the positions of the $R_0$ and $T_0$ conditions to the linewidth of a resonance. We consider a resonance of the odd part of the response, with complex frequency $\tilde{\omega}=\omega_\text{r}+j\Gamma$, as defined in Sec.~\ref{sec_model}, where $\Gamma>0$, while $s_\text{e}$ does not resonate and instead varies slowly. As shown in Appendix~\ref{app_S}, $|s_\text{o}|=1$ for all real frequencies for a lossless metasurface. Since $s_\text{o}$ resonates, it has a pole at $\tilde{\omega}$, and therefore also a zero at $\tilde{\omega}^*=\omega_\text{r}-j\Gamma$. Indeed, to ensure that $|s_\text{o}|=1$, the zero and the pole of $s_\text{o}$ must be at the same distance from every point of the real frequency axis, which is only possible if the zero is the mirror image of the pole with respect to this axis, i.e., if it is located at $\tilde{\omega}^*$.

Close to the resonance, we can thus write~\cite{grigoriev2013} 
\begin{equation}
\label{eq_so_res}
s_\text{o}\approx s_\text{bg}\,\frac{\omega-\omega_\text{r}+j\Gamma}{\omega-\omega_\text{r}-j\Gamma},
\end{equation}
where $s_\text{bg}$ is the background, i.e., the value of $s_\text{o}$ away from the resonance, as in~\eqref{eq_dw_RT}.

To obtain the phase of the resonant factor in~\eqref{eq_so_res}, we write $x=\omega-\omega_\text{r}$, which is real for real frequencies, and multiply the numerator and the denominator of this factor by $x+j\Gamma$. This gives
\begin{equation}
\label{eq_F}
\begin{split}
\frac{x+j\Gamma}{x-j\Gamma}&=\frac{(x+j\Gamma)^2}{x^2+\Gamma^2}\\
&=\frac{x^2-\Gamma^2}{x^2+\Gamma^2}+j\,\frac{2x\Gamma}{x^2+\Gamma^2}.
\end{split}
\end{equation}
The sum of the squares of the real and imaginary parts of~\eqref{eq_F} is equal to one. The resonant factor thus has unit magnitude and can be written as $e^{j\varphi}$, with
\begin{equation}
\label{eq_cossin}
\cos\varphi=\frac{x^2-\Gamma^2}{x^2+\Gamma^2},\qquad\sin\varphi=\frac{2x\Gamma}{x^2+\Gamma^2}.
\end{equation}
Using the identity $\tan(\varphi/2)=\sin\varphi/(1+\cos\varphi)$, we find $\tan(\varphi/2)=\Gamma/x$, so that, modulo $2\pi$,
\begin{equation}
\label{eq_phi_res}
\varphi=2\arctan\left(\frac{\Gamma}{\omega-\omega_\text{r}}\right).
\end{equation}
From~\eqref{eq_phi_res}, we see that as $\omega$ goes from far below to far above $\omega_\text{r}$, its phase changes by $2\pi$, and half of it occurs within the linewidth of the resonance between $\omega_\text{r}-\Gamma$ and $\omega_\text{r}+\Gamma$. 

Now, since the metasurface is lossless, we have that $s_\text{bg}=e^{j\varphi_\text{bg}}$, which allows us to express~\eqref{eq_so_res} as
\begin{equation}
\label{eq_so_phase}
s_\text{o}\approx e^{j\left(\varphi_\text{bg}+\varphi\right)}.
\end{equation}
As in~\eqref{eq_dw_RT}, we denote the phase difference between $s_\text{bg}$ and $s_\text{e}$ as $\Delta\varphi=\varphi_\text{bg}-\varphi_\text{e}$, where $s_\text{e}=e^{j\varphi_\text{e}}$. Combining these expressions with~\eqref{eq_so_phase}, we have
\begin{equation}
\frac{s_\text{o}}{s_\text{e}}\approx e^{j\left(\varphi+\Delta\varphi\right)}.
\end{equation}
The $R_0$ and $T_0$ conditions, $s_\text{o}=s_\text{e}$ and $s_\text{o}=-s_\text{e}$, thus correspond to $e^{j(\varphi+\Delta\varphi)}=1$ and $e^{j(\varphi+\Delta\varphi)}=-1$, respectively, which gives
\begin{subequations}
\begin{align}
R_0:&\quad\varphi=-\Delta\varphi,\\
T_0:&\quad\varphi=\pi-\Delta\varphi,
\end{align}
\end{subequations}
modulo $2\pi$. Using~\eqref{eq_phi_res}, the frequencies of these two conditions are
\begin{subequations}
\label{eq_wRT}
\begin{align}
\omega_{R_0}-\omega_\text{r}&=-\Gamma\cot\left(\frac{\Delta\varphi}{2}\right),\\
\omega_{T_0}-\omega_\text{r}&=\Gamma\tan\left(\frac{\Delta\varphi}{2}\right).
\end{align}
\end{subequations}
Both conditions are thus located within a few linewidths of the resonance, unless $\Delta\varphi$ is close to $0$ or $\pi$. Their separation is
\begin{equation}
\begin{split}
|\omega_{R_0}-\omega_{T_0}|&=\Gamma\left|\cot\left(\frac{\Delta\varphi}{2}\right)+\tan\left(\frac{\Delta\varphi}{2}\right)\right|\\
&=\frac{2\Gamma}{|\sin\Delta\varphi|},
\end{split}
\end{equation}
which is~\eqref{eq_dw_RT}.

For $\Delta\varphi\approx 0$, \eqref{eq_wRT} shows that the $T_0$ condition lies at the center of the resonance, while the $R_0$ condition is pushed away from it, on a side given by the sign of $\Delta\varphi$. The opposite occurs for $\Delta\varphi\approx\pi$. Moreover, multiplying the two expressions in~\eqref{eq_wRT}, we find
\begin{equation}
\Gamma^2=|\omega_{R_0}-\omega_\text{r}|\,|\omega_{T_0}-\omega_\text{r}|,
\end{equation}
which implies that the half-linewidth of the resonance, $\Gamma$, is the geometric mean of the distances between the two zeros and $\omega_\text{r}$. The $R_0$ and $T_0$ conditions also lie on the circle that passes through the pole and the zero and is centered on the real axis at $\omega_\text{r}-\Gamma\cot\Delta\varphi$. Indeed, this circle has a radius $\Gamma/|\sin\Delta\varphi|$, and it intersects the real axis at $\omega_\text{r}-\Gamma\cot\Delta\varphi\pm\Gamma/|\sin\Delta\varphi|$, which coincide with~\eqref{eq_wRT}. The diameter of this circle is thus equal to the separation in~\eqref{eq_dw_RT}.

These results can also be related to the Fano lineshape of the resonance~\cite{fan2003}. From~\eqref{eq_RTeo} and the expressions above, we have that
\begin{subequations}
\begin{align}
T &= +\frac{s_\text{e}}{2}\left[1+e^{j(\varphi+\Delta\varphi)}\right],\\  
R &= -\frac{s_\text{e}}{2}\left[1-e^{j(\varphi+\Delta\varphi)}\right],
\end{align}
\end{subequations}
which leads to
\begin{subequations}
\begin{align}
|T|^2&=\cos^2\left(\frac{\varphi+\Delta\varphi}{2}\right),\\
|R|^2&=\sin^2\left(\frac{\varphi+\Delta\varphi}{2}\right).
\end{align}
\end{subequations}
Using~\eqref{eq_phi_res} and defining the normalized frequency $\varepsilon=(\omega-\omega_\text{r})/\Gamma$, the transmittance becomes
\begin{equation}
\label{eq_fano}
\begin{gathered}
|T|^2=\cos^2\left(\frac{\Delta\varphi}{2}\right)\frac{(\varepsilon+q)^2}{\varepsilon^2+1},\\
q=-\tan\left(\frac{\Delta\varphi}{2}\right),
\end{gathered}
\end{equation}
which corresponds to the well-known Fano lineshape with the asymmetry parameter $q$.~\cite{limonov2017} Similarly, $|R|^2$ is obtained from~\eqref{eq_fano} by replacing $\cos^2(\Delta\varphi/2)$ by $\sin^2(\Delta\varphi/2)$ and $q$ by $\cot(\Delta\varphi/2)$. For $\Delta\varphi=0$, we have $q=0$, and $|T|^2$ is a symmetric dip, while for $\Delta\varphi=\pi$, we have $|q|\rightarrow\infty$, and $|T|^2$ is a symmetric Lorentzian peak. For other values of $\Delta\varphi$, the lineshape is asymmetric. Note that the zero and the maximum of~\eqref{eq_fano}, located at $\varepsilon=-q$ and $\varepsilon=1/q$, respectively, correspond to the $T_0$ and $R_0$ conditions in~\eqref{eq_wRT}.

Finally, the same results hold for a resonance of $s_\text{e}$, with the roles of $s_\text{e}$ and $s_\text{o}$ exchanged.

\section{Resonance Frequencies and Q Factors}
\label{app_Q}

In this appendix, we first relate the imaginary part, $\Gamma$, of the resonance frequency to the coupling coefficient of a resonance, which is used in Secs.~\ref{sec_examples} and~\ref{sec_passive}. We then present a general procedure to obtain the resonance frequencies and Q factors from the poles of $s_\text{e}$ and $s_\text{o}$, and apply it to the metasurfaces considered in this work.

\subsection{Radiation Rate and Coupling Coefficient}
\label{app_rad}

We derive~\eqref{eq_ImWd} using temporal coupled-mode theory~\cite{fan2003}. We consider a single resonance with amplitude $a$, normalized so that $|a|^2$ is the energy stored in the resonance. The waves incident on and radiated by the metasurface have amplitudes $s_+$ and $s_-$, respectively, normalized so that $|s_\pm|^2$ is the power that they carry. The resonance and the radiated wave are then described by
\begin{subequations}
\label{eq_tcmt}
\begin{align}
\frac{\mathrm{d}a}{\mathrm{d}t}&=j\tilde{\omega}\,a+\kappa\,s_+,\\
s_-&=s_\text{bg}\,s_++d\,a,
\end{align}
\end{subequations}
where $\tilde{\omega}$ is the complex frequency of the resonance, $\kappa$ and $d$ are the coupling coefficients of the resonance to the incident and radiated waves, respectively, and $s_\text{bg}$ describes the non-resonant scattering. Without incident wave, i.e., for $s_+=0$, the resonance decays as $a\propto e^{j\tilde{\omega}t}=e^{j\omega_\text{r}t}\,e^{-\Gamma t}$. The stored energy then decreases as
\begin{equation}
\label{eq_2gamma} 
\frac{\mathrm{d}|a|^2}{\mathrm{d}t}=-2\Gamma\,|a|^2.
\end{equation}
Since the metasurface is lossless, this energy can only leave the resonance as radiation, whose power is
\begin{equation}
    \label{eq_scP} 
|s_-|^2=|d|^2|a|^2.
\end{equation}
Equating the energy lost by the resonance from~\eqref{eq_2gamma} with the radiated power in~\eqref{eq_scP}, we obtain $2\Gamma=|d|^2$, which is~\eqref{eq_ImWd}. Moreover, reciprocity implies that $\kappa=d$~\cite{fan2003}, i.e., the resonance couples to the incident wave exactly as strongly as it radiates. Since $|d|^2\geq 0$, $\Gamma$ can never be negative for a passive resonance.

\subsection{General Procedure}
\label{app_proc}

As discussed in Sec.~\ref{sec_model}, the resonances of the metasurface correspond to the poles of $s_\text{e}$ and $s_\text{o}$ in~\eqref{eq_seso}. They are thus given by the complex frequencies $\tilde{\omega}$ at which
\begin{subequations}
\label{eq_poles}
\begin{align}
&2+jk_z\chi_\text{ee}^{xx}=0,\label{eq_pole_e}\\
&2k_z+j\left(k^2\chi_\text{mm}^{yy}+k_x^2\chi_\text{ee}^{zz}\right)=0,\label{eq_pole_o}
\end{align}
\end{subequations}
for the even and odd parts of the response, respectively.

We consider a Lorentzian resonance with $\gamma=0$ in~\eqref{eq_lorentz} and write the complex resonance frequency as
\begin{equation}
\label{eq_dw}
\tilde{\omega}=\omega_0+\delta\omega,
\end{equation}
with $|\delta\omega|\ll\omega_0$. Close to the resonance, the denominator of~\eqref{eq_lorentz}, evaluated at $\omega=\tilde{\omega}$, can then be approximated as $\omega_0^2-\tilde{\omega}^2\approx-2\omega_0\delta\omega$, so that the susceptibility becomes
\begin{equation}
\label{eq_chi_dw}
\chi_\text{aa}^{bb}\approx-\frac{A_\text{aa}^{bb}}{2\omega_0\delta\omega}.
\end{equation}
Substituting~\eqref{eq_chi_dw} into~\eqref{eq_poles} leads to a simple linear equation for $\delta\omega$. Since $\omega_0$ is real, we have $\Gamma=\text{Im}\{\delta\omega\}$, while $\omega_\text{r}=\omega_0+\text{Re}\{\delta\omega\}\approx\omega_0$. The Q factor is then obtained from
\begin{equation}
\label{eq_Qdef}
Q=\frac{\omega_\text{r}}{2\Gamma}\approx\frac{\omega_0}{2\,\text{Im}\{\delta\omega\}}.
\end{equation}
In the following, we apply this procedure to the cases discussed in the main text.

\subsection{Symmetry-Protected BIC}
\label{app_sym}

We first consider the metasurface of Fig.~\ref{fig_PM_AmpPhase}, discussed in~\subsecref{sec_MyPz}, for which $\chi_\text{ee}^{xx}=0$. We derive the Q factor of the symmetry-protected BIC, then that of the sharp transmission feature, and finally that of the quasi-BIC obtained by breaking the in-plane symmetry of the particles.

\subsubsection{Q Factor Near Normal Incidence}
\label{app_sym_normal}

We consider the resonance of $\chi_\text{ee}^{zz}$. We assume that $\chi_\text{mm}^{yy}$ varies slowly in the vicinity of this resonance, so that it can simply be evaluated at $\omega_0$. Substituting~\eqref{eq_chi_dw} for $\chi_\text{ee}^{zz}$ into~\eqref{eq_pole_o} gives
\begin{equation}
2k_z+jk^2\chi_\text{mm}^{yy}-j\frac{k_x^2A_\text{ee}^{zz}}{2\omega_0\delta\omega}=0,
\end{equation}
whose solution is
\begin{equation}
\delta\omega=\frac{jk_x^2A_\text{ee}^{zz}}{2\omega_0\left(2k_z+jk^2\chi_\text{mm}^{yy}\right)}.
\end{equation}
Taking the imaginary part of this expression, we find
\begin{equation}
\label{eq_ImWzz}
    \Gamma = \text{Im}\{\delta \omega\}\approx\frac{k_zk_x^2A_\text{ee}^{zz}}{\omega_0\left[4k_z^2+k^4\left(\chi_\text{mm}^{yy}\right)^2\right]},
\end{equation}
where all quantities are evaluated at $\omega_0$. Since $k_x=k\sin\theta$, $\Gamma$ is proportional to $\sin^2\theta$. From~\eqref{eq_Qdef}, this leads to~\eqref{eq_Qsin} at small angles, with a Q factor that diverges at normal incidence.

\subsubsection{Transmission Peak}
\label{app_sym_peak}

Since $\chi_\text{ee}^{xx}=0$, we have $s_\text{e}=1$. From~\eqref{eq_RTeo} and~\eqref{eq_seso}, the transmission coefficient then reduces to
\begin{equation}
\label{eq_TY}
T=\frac{1+s_\text{o}}{2}=\frac{2k_z}{2k_z+jY},
\end{equation}
where we have defined
\begin{equation}
\label{eq_Y}
Y=k^2\chi_\text{mm}^{yy}+k_x^2\chi_\text{ee}^{zz}.
\end{equation}
At a given incidence angle, full transmission is obtained at the frequency $\omega_\text{p}$ for which $Y=0$, which corresponds to the $R_0^\text{inv}$ condition in~\eqref{eq_R0p}.

The Q factor of the corresponding resonance can be obtained from the pole of $s_\text{o}$. Since the peak does not coincide with the resonance of $\chi_\text{ee}^{zz}$ or $\chi_\text{mm}^{yy}$, we cannot use~\eqref{eq_chi_dw} here. Instead, since $Y$ varies smoothly around $\omega_\text{p}$, we expand it in a Taylor series around this frequency, i.e.,
\begin{equation}
\label{eq_Ytaylor}
Y(\omega)=Y(\omega_\text{p})+(\omega-\omega_\text{p})Y'+\dots,
\end{equation}
where $Y'=\partial Y/\partial\omega$ is evaluated at $\omega_\text{p}$. By definition of $\omega_\text{p}$, we have $Y(\omega_\text{p})=0$, which leads to
\begin{equation}
\label{eq_Yw}
Y\approx (\omega-\omega_\text{p})Y'.
\end{equation}
Since $k_x=k\sin\theta$, we can write $Y=k^2(\chi_\text{mm}^{yy}+\sin^2\theta\,\chi_\text{ee}^{zz})$. At $\omega_\text{p}$, the term in parentheses vanishes, so that only the frequency derivatives of the susceptibilities contribute to $Y'$. Using~\eqref{eq_lorentz} with $\gamma=0$, for which $\partial\chi_\text{aa}^{bb}/\partial\omega=2A_\text{aa}^{bb}\omega/(\omega_0^2-\omega^2)^2$, we then obtain
\begin{equation}
\label{eq_Yp}
Y'=2k_\text{p}^2\,\omega_\text{p}\,\Sigma,
\end{equation}
with $k_\text{p}=\omega_\text{p}/c$ and
\begin{equation}
\label{eq_Sigma}
\Sigma=\frac{A_\text{mm}^{yy}}{\left[\left(\omega_\text{mm}^{yy}\right)^2-\omega_\text{p}^2\right]^2}
+\frac{\sin^2\theta A_\text{ee}^{zz}}{\left[\left(\omega_\text{ee}^{zz}\right)^2-\omega_\text{p}^2\right]^2},
\end{equation}
where $\omega_\text{mm}^{yy}$ and $\omega_\text{ee}^{zz}$ are the resonance frequencies of $\chi_\text{mm}^{yy}$ and $\chi_\text{ee}^{zz}$, respectively. Note that $Y'>0$ when both amplitudes are positive, since each term of $\Sigma$ is then positive. The pole of $s_\text{o}$, and thus of $T$, is located at the complex frequency $\tilde{\omega}$ for which the denominator of~\eqref{eq_TY} vanishes, i.e., $2k_z+jY(\tilde{\omega})=0$. Substituting~\eqref{eq_Yw} into this condition, and evaluating $k_z$ at $\omega_\text{p}$, which is valid as long as $|\tilde{\omega}-\omega_\text{p}|\ll\omega_\text{p}$, we obtain
\begin{equation}
2k_z+j(\tilde{\omega}-\omega_\text{p})Y'=0,
\end{equation}
whose solution is
\begin{equation}
\label{eq_pole_p}
\tilde{\omega}\approx\omega_\text{p}+j\frac{2k_z}{Y'}.
\end{equation}
Since $k_z$ and $Y'$ are both real, the correction to $\omega_\text{p}$ is purely imaginary. We thus have $\omega_\text{r}\approx\omega_\text{p}$ and $\Gamma\approx 2k_z/Y'$, which is positive since $Y'>0$, and the Q factor reads
\begin{equation}
\label{eq_Qspec}
Q\approx\frac{\omega_\text{p}Y'}{4k_z}.
\end{equation}
Using~\eqref{eq_Yp} and $k_z=k_\text{p}\cos\theta$, this becomes
\begin{equation}
\label{eq_Qspec2}
Q\approx\frac{\omega_\text{p}^2\,k_\text{p}}{2\cos\theta}\,\Sigma.
\end{equation}
Increasing the amplitudes of both $\chi_\text{mm}^{yy}$ and $\chi_\text{ee}^{zz}$ by a common factor does not change $\omega_\text{p}$, since the condition $Y=0$ is unchanged, but increases $\Sigma$, and thus $Q$, by the same factor. Finally, \eqref{eq_Qspec2} predicts a diverging Q factor near the two tips of the transmission feature. At grazing incidence, this is due to the factor $1/\cos\theta$. At normal incidence, the peak approaches the resonance of $\chi_\text{ee}^{zz}$, where the second term of $\Sigma$ diverges. There, however, the linear expansion in~\eqref{eq_Yw} is no longer accurate and~\eqref{eq_ImWzz} should be used instead.

\subsubsection{Quasi-BIC from In-Plane Asymmetry}
\label{app_sym_qbic}

We now add to $\chi_\text{mm}^{yy}$ a Lorentzian resonance with amplitude $A_\text{mm}^{yy,\alpha}$ at the resonance frequency $\omega_0$ of $\chi_\text{ee}^{zz}$. Using~\eqref{eq_chi_dw} for both resonant terms, the term in parentheses in~\eqref{eq_pole_o} becomes
\begin{equation}
k^2\chi_\text{mm}^{yy}+k_x^2\chi_\text{ee}^{zz}\approx k^2\chi_\text{mm}^{yy}-\frac{k^2A_\text{mm}^{yy,\alpha}+k_x^2A_\text{ee}^{zz}}{2\omega_0\delta\omega},
\end{equation}
This expression has the same form as in the symmetry-protected case, with $k_x^2A_\text{ee}^{zz}$ replaced by $k^2A_\text{mm}^{yy,\alpha}+k_x^2A_\text{ee}^{zz}$. Following the same steps as for~\eqref{eq_ImWzz}, we thus find
\begin{equation}
\label{eq_ImWasym}
    \Gamma\approx\frac{k_z\left(k^2A_\text{mm}^{yy,\alpha}+k_x^2A_\text{ee}^{zz}\right)}{\omega_0\left[4k_z^2+k^4\left( \chi_\text{mm}^{yy} \right)^2\right]}.
\end{equation}
At normal incidence, $k_x=0$ and $\Gamma$ is proportional to $A_\text{mm}^{yy,\alpha}$. From~\eqref{eq_Qdef}, the Q factor is thus inversely proportional to $A_\text{mm}^{yy,\alpha}$, which leads to~\eqref{eq_Qalpha}. Moreover, since both amplitudes are positive, \eqref{eq_ImWasym} does not vanish at any angle, which confirms that the BIC has become a quasi-BIC.

\subsection{Tuning the Q Factor with $\chi_\text{ee}^{xx}$}
\label{app_xxQ}

For the metasurface of Fig.~\ref{fig_AC}, discussed in~\subsecref{sec_xx}, we consider the resonance of the renormalized susceptibility $\chi_\text{ee}^{'xx}=\chi_\text{ee}^{xx}/C$. From~\eqref{eq_chi_dw}, we have
\begin{equation}
\chi_\text{ee}^{'xx}\approx-\frac{A_\text{ee}^{xx}}{2C\omega_0\delta\omega}.
\end{equation}
Substituting this expression into~\eqref{eq_pole_e} gives
\begin{equation}
2-j\frac{k_zA_\text{ee}^{xx}}{2C\omega_0\delta\omega}=0,
\end{equation}
whose solution is
\begin{equation}
\delta\omega=j\frac{k_zA_\text{ee}^{xx}}{4C\omega_0}.
\end{equation}
Using~\eqref{eq_Qdef}, this directly leads to the Q factor in~\eqref{eq_Qxx}.

\subsection{Accidental BIC}
\label{app_acc}

For the metasurface of Fig.~\ref{fig_C}, discussed in~\subsecref{sec_accidental}, $\chi_\text{ee}^{zz}$ and $\chi_\text{mm}^{yy}$ share the same resonance frequency $\omega_0$. Using~\eqref{eq_chi_dw} for both susceptibilities, the term in parentheses in~\eqref{eq_pole_o} becomes
\begin{equation}
\label{eq_shared}
k^2\chi_\text{mm}^{yy}+k_x^2\chi_\text{ee}^{zz}\approx-\frac{k^2\left(A_\text{mm}^{yy}+A_\text{ee}^{zz}\sin^2\theta\right)}{2\omega_0\delta\omega}.
\end{equation}
Substituting~\eqref{eq_shared} into~\eqref{eq_pole_o} gives
\begin{equation}
\delta\omega=j\frac{k^2\left(A_\text{mm}^{yy}+A_\text{ee}^{zz}\sin^2\theta\right)}{4\omega_0k_z},
\end{equation}
which is purely imaginary and directly leads to~\eqref{eq_ImW}. Using~\eqref{eq_Qdef}, the corresponding Q factor is
\begin{equation}
\label{eq_Qacc}
Q\approx\frac{2\omega_0^2k_z}{k^2\left(A_\text{mm}^{yy}+A_\text{ee}^{zz}\sin^2\theta\right)}.
\end{equation}
This Q factor diverges at the BIC angle, $\theta_\text{BIC}$, which is defined by $A_\text{mm}^{yy}+A_\text{ee}^{zz}\sin^2\theta_\text{BIC}=0$, in agreement with~\eqref{eq_crossing}. To obtain its behavior in the vicinity of this angle, we write $\theta=\theta_\text{BIC}+\Delta\theta$, with $|\Delta\theta|\ll 1$, and expand $\sin^2\theta$ to first order in $\Delta\theta$, which gives
\begin{equation}
\label{eq_sinexp_bic}
\sin^2\theta\approx\sin^2\theta_\text{BIC}+\sin(2\theta_\text{BIC})\,\Delta\theta.
\end{equation}
Using the definition of $\theta_\text{BIC}$, the term in parentheses in~\eqref{eq_Qacc} then becomes
\begin{equation}
\label{eq_lin}
A_\text{mm}^{yy}+A_\text{ee}^{zz}\sin^2\theta\approx A_\text{ee}^{zz}\sin(2\theta_\text{BIC})\,\Delta\theta.
\end{equation}
Substituting~\eqref{eq_lin} into~\eqref{eq_Qacc}, we finally obtain
\begin{equation}
\label{eq_QBIC}
Q\approx\frac{2\omega_0^2k_z}{k^2A_\text{ee}^{zz}\sin(2\theta_\text{BIC})\,\Delta\theta},
\end{equation}
which shows that $|Q|\propto|\Delta\theta|^{-1}=|\theta-\theta_\text{BIC}|^{-1}$. Since $A_\text{ee}^{zz}<0$, the Q factor is positive for $\theta<\theta_\text{BIC}$ and negative for $\theta>\theta_\text{BIC}$. A negative Q factor corresponds to $\Gamma<0$, i.e., to a resonance that is growing in time, as discussed in~\subsecref{sec_accidental}.

\subsection{Passive Model of an Accidental BIC}
\label{app_passiveQ}

For the passive nonlocal model of Sec.~\ref{sec_passive}, we substitute~\eqref{eq_passive_chi} into the term in parentheses in~\eqref{eq_pole_o}. Using $k_x=k\sin\theta$, this term becomes
\begin{equation}
k^2\chi_\text{mm}^{yy}+k_x^2\chi_\text{ee}^{zz}=\frac{k^2W(\theta)}{\Omega^2-\omega^2},
\end{equation}
where $W(\theta)=(a+b\sin^2\theta)^2$ is given by~\eqref{eq_square}. In contrast with the previous cases, the resonance frequency $\Omega$ depends on $k_x$, which itself depends on the frequency, since $k_x=(\omega/c)\sin\theta$ at a fixed angle of incidence. We thus have $c^2k_x^2=\omega^2\sin^2\theta$, so that $\Omega^2=\omega_0^2-\eta\,\omega^2\sin^2\theta$ and the denominator of the Lorentzian becomes
\begin{equation}
\label{eq_Omega_fixed}
\Omega^2-\omega^2=\omega_0^2-\left(1+\eta\sin^2\theta\right)\omega^2.
\end{equation}
This denominator vanishes at
\begin{equation}
\label{eq_wr_passive}
\omega_\text{r}=\frac{\omega_0}{\sqrt{1+\eta\sin^2\theta}},
\end{equation}
which is the resonance frequency at the angle $\theta$. For $\eta>0$, it decreases as the angle increases, which corresponds to the red shift of the resonance shown in Fig.~\ref{fig_passive}. Using~\eqref{eq_wr_passive}, we can rewrite~\eqref{eq_Omega_fixed} as $\Omega^2-\omega^2=(1+\eta\sin^2\theta)(\omega_\text{r}^2-\omega^2)$. We then proceed as in the general procedure above, by writing the complex resonance frequency as $\tilde{\omega}=\omega_\text{r}+\delta\omega$, with $|\delta\omega|\ll\omega_\text{r}$, so that
\begin{equation}
\Omega^2-\tilde{\omega}^2\approx -2\left(1+\eta\sin^2\theta\right)\omega_\text{r}\,\delta\omega.
\end{equation}
Substituting these expressions into~\eqref{eq_pole_o} gives
\begin{equation}
2k_z-j\frac{k^2W(\theta)}{2\left(1+\eta\sin^2\theta\right)\omega_\text{r}\,\delta\omega}=0,
\end{equation}
whose solution is
\begin{equation}
\label{eq_dw_passive}
\delta\omega=j\frac{k^2\left(a+b\sin^2\theta\right)^2}{4\left(1+\eta\sin^2\theta\right)\omega_\text{r}k_z},
\end{equation}
where $k$ and $k_z$ are evaluated at $\omega_\text{r}$. As for the accidental BIC of the local dipolar model in~\eqref{eq_ImW}, $\delta\omega$ is purely imaginary and $\Gamma=\text{Im}\{\delta\omega\}$. For $\eta=0$, we have $\omega_\text{r}=\omega_0$, and~\eqref{eq_dw_passive} reduces to the same form as~\eqref{eq_ImW}, with $A_\text{mm}^{yy}+A_\text{ee}^{zz}\sin^2\theta$ replaced by $(a+b\sin^2\theta)^2$. In contrast with the local dipolar model, however, $\Gamma$ is never negative, since all the factors in~\eqref{eq_dw_passive} are positive, and it only vanishes at $\theta_\text{BIC}$, where $a+b\sin^2\theta_\text{BIC}=0$.

To obtain the behavior of the Q factor close to the BIC, we write $\theta=\theta_\text{BIC}+\Delta\theta$ and use~\eqref{eq_sinexp_bic}, which gives $a+b\sin^2\theta\approx b\sin(2\theta_\text{BIC})\,\Delta\theta$. From~\eqref{eq_Qdef}, we then obtain
\begin{equation}
\label{eq_Q_passive}
Q=\frac{\omega_\text{r}}{2\Gamma}\approx\frac{2\left(1+\eta\sin^2\theta_\text{BIC}\right)\omega_\text{r}^2k_z}{k^2b^2\sin^2(2\theta_\text{BIC})\,\Delta\theta^2},
\end{equation}
which shows that $Q\propto(\theta-\theta_\text{BIC})^{-2}$, with a positive Q factor on both sides of the BIC, in contrast with~\eqref{eq_QBIC}.

\subsection{Parameters of the Silicon-Ridge Metasurface}
\label{app_fit}

\begin{figure}[t]
	\centering
	\includegraphics[width=\linewidth]{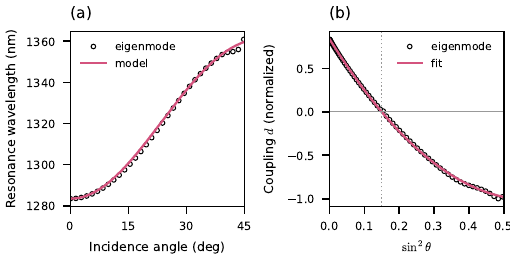}
	\caption{Extraction of the parameters of the passive nonlocal model for the silicon-ridge metasurface. (a) Resonance wavelength of the eigenmode (circles) and of the model (solid). (b) Coupling coefficient extracted from the linewidth of the eigenmode (circles), normalized by its maximum, and its fit by $d=(a+b\sin^2\theta)\cos^2\theta$ (solid). The dotted line indicates $\theta_\text{BIC}=22.7^\circ$.}
	\label{fig_fit}
\end{figure}

The parameters of the nonlocal passive model of Sec.~\ref{sec_passive} are extracted from the complex frequency $\tilde{\omega}=\omega_\text{r}+j\Gamma$ of the resonant mode of the silicon-ridge metasurface, obtained from the eigenmode analysis of~\subsecref{sec_ridge}. For the angular dispersion, we consider the next order in $\sin^2\theta$ in the resonance frequency of~\eqref{eq_passive_chi}, i.e., $\Omega^2=\omega_0^2-(\eta\sin^2\theta+\zeta\sin^4\theta)\,\omega^2$, so that $\omega_\text{r}=\omega_0/\sqrt{1+\eta\sin^2\theta+\zeta\sin^4\theta}$ at a fixed angle. As shown in Fig.~\ref{fig_fit}a, fitting the eigen-wavelengths gives $\lambda_0=1283.4$~nm, $\eta=0.429$ and $\zeta=-0.367$. Following Appendix~\ref{app_passiveQ}, with $1+\eta\sin^2\theta$ replaced by $1+\eta\sin^2\theta+\zeta\sin^4\theta$, the imaginary part of the resonance frequency is
\begin{equation}
\label{eq_Gamma_ridge}
\Gamma\approx\frac{k^2W(\theta)}{4\left(1+\eta\sin^2\theta+\zeta\sin^4\theta\right)\omega_\text{r}k_z},
\end{equation}
where $k$ and $k_z$ are evaluated at $\omega_\text{r}$. This allows us to extract $W(\theta)$, and thus $d=\pm\sqrt{W(\theta)}$, from the linewidth of the eigenmode, with a sign change at the BIC. As shown in Fig.~\ref{fig_fit}b, $d$ is accurately described, with a relative error of about $1\%$, by $d=(a+b\sin^2\theta)\cos^2\theta$, with the values of $a$ and $b$ given in Table~\ref{tab_params}. The coupling coefficient thus vanishes for $\sin^2\theta_\text{BIC}=-a/b$, i.e., $\theta_\text{BIC}=22.7^\circ$.

The susceptibilities of the model are then constructed as follows. Their resonant parts are lossless Lorentzian functions sharing the denominator $\Omega^2-\omega^2$, with the amplitude $a^2$ for $\chi_\text{mm}^{yy}$ and the angle-dependent amplitude $[W(\theta)-a^2]/\sin^2\theta$ for $\chi_\text{ee}^{zz}$, so that the term $k^2\chi_\text{mm}^{yy}+k_x^2\chi_\text{ee}^{zz}$ in~\eqref{eq_seso} contains the resonant contribution $k^2W(\theta)/(\Omega^2-\omega^2)$, with $W(\theta)=d^2(\theta)$. All the parameters of these resonant parts are thus determined by the eigenmode analysis. We then add to $\chi_\text{ee}^{xx}$, $\chi_\text{mm}^{yy}$ and $\chi_\text{ee}^{zz}$ a non-resonant background contribution, which accounts for the resonances located outside of the considered wavelength range, and which we take as a linear function of frequency, $c_0+c_1(\omega/\omega_\text{c}-1)$, where $\omega_\text{c}$ corresponds to the wavelength $\lambda=1325$~nm. The six coefficients of these backgrounds are obtained by a least-squares fit of the transmittance computed from~\eqref{eq_RTeo} and~\eqref{eq_seso} to the full-wave transmittance, over the wavelength range from $1200$ to $1450$~nm and the angular range from $0^\circ$ to $45^\circ$, while the resonant parts are kept fixed. All the parameters of the model, including the fitted coefficients of the non-resonant contributions, are given in Table~\ref{tab_params}.

\begin{table}[htb]
\caption{Parameters of the passive model of the silicon-ridge metasurface. The resonant part is extracted from the eigenmode analysis, while the non-resonant contributions, $c_0+c_1(\omega/\omega_\text{c}-1)$ with $\omega_\text{c}=2\pi c/(1325~\text{nm})$, are fitted to the full-wave transmittance.}
\label{tab_params}
\centering
\begin{tabular}{lcc}
\hline\hline
\multicolumn{3}{l}{Resonant part}\\
\hline
$\lambda_0$ & \multicolumn{2}{c}{$1283.4$~nm}\\
$\eta$ & \multicolumn{2}{c}{$0.429$}\\
$\zeta$ & \multicolumn{2}{c}{$-0.367$}\\
$a$ & \multicolumn{2}{c}{$1.317\times10^{11}$~m$^{1/2}\cdot$s$^{-1}$}\\
$b$ & \multicolumn{2}{c}{$-8.815\times10^{11}$~m$^{1/2}\cdot$s$^{-1}$}\\
\hline
Non-resonant part & $c_0$ ($10^{-7}$~m) & $c_1$ ($10^{-7}$~m)\\
\hline
$\chi_\text{ee}^{xx}$ & $0.359$ & $-34.53$\\
$\chi_\text{mm}^{yy}$ & $-0.976$ & $-3.172$\\
$\chi_\text{ee}^{zz}$ & $6.579$ & $-41.23$\\
\hline\hline
\end{tabular}
\end{table}

\section{Double Avoided Crossing}
\label{app_DC}

\begin{figure}[b]
	\centering
	\includegraphics[width=\linewidth]{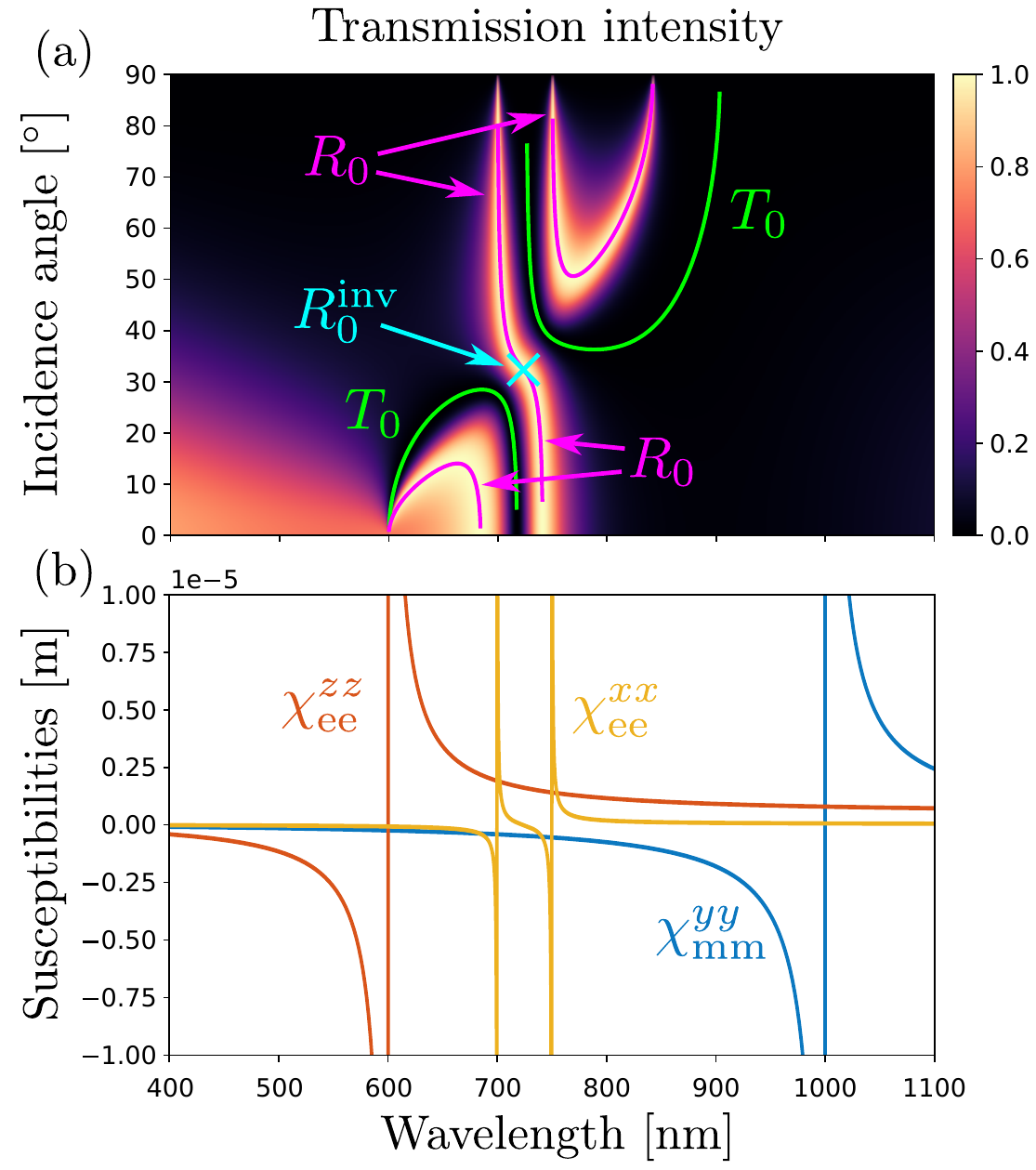}
	\caption{Angular scattering response of a metasurface with a double $\chi_\text{ee}^{xx}$ resonance. The susceptibilities $\chi_\text{ee}^{zz}$ and $\chi_\text{mm}^{yy}$ are the same as in Fig.~\ref{fig_PM_AmpPhase}, i.e., $A_\text{ee}^{zz}=5\times 10^{24}$, $\lambda_\text{ee}^{zz}=600$~nm, $A_\text{mm}^{yy}=1.5\times 10^{24}$ and $\lambda_\text{mm}^{yy}=1000$~nm, while $\chi_\text{ee}^{xx}$ is the sum of two Lorentzian resonances with $A_\text{ee,1}^{xx}=A_\text{ee,2}^{xx}=10^{23}$, $\lambda_{01,\text{ee}}^{xx}=700$~nm and $\lambda_{02,\text{ee}}^{xx}=750$~nm, so that $\chi_\text{ee}^{xx}=0$ at $\lambda\approx 724$~nm. (a) Transmittance and (b) real part of the susceptibilities. The pink and green solid lines denote the $R_{0}$ and $T_{0}$ conditions, respectively. The $R_0^\text{inv}$ condition is therefore only satisfied at a single point, at this wavelength and $\theta\approx 32^\circ$.}
	\label{fig_DC}
\end{figure}

In this appendix, we investigate what happens if an $R_0^\text{inv}$ condition is forced in the middle of the avoided-crossing region of Fig.~\ref{fig_AC}c. We recall from~\eqref{eq_R0p} that the $R_0^\text{inv}$ condition can only exist if $\chi_\text{ee}^{xx} = 0$ and if the ratio $-\chi_\text{mm}^{yy}/\chi_\text{ee}^{zz}$ leads to a real-valued angle. We already know that this susceptibility ratio can lead to an $R_0^\text{inv}$ condition in the wavelength range of the avoided crossing region since this corresponds to what we had in Fig.~\ref{fig_PM_AmpPhase}. Now, to simultaneously obtain $\chi_\text{ee}^{xx} = 0$ in that wavelength range while being nonzero elsewhere, we express $\chi_\text{ee}^{xx}$ in terms of \emph{two} Lorentzian resonances with equal amplitudes, located at 700~nm and 750~nm, such that $\chi_\text{ee}^{xx}=0$ at $\lambda \approx 724$~nm. The resulting transmittance map and corresponding susceptibilities are plotted in Fig.~\ref{fig_DC}.

Comparing the transmittance maps of Fig.~\ref{fig_AC} and Fig.~\ref{fig_DC}, we see that forcing the existence of an $R_0^\text{inv}$ condition leads to a double avoided crossing. This time, it is the $T_0$ condition that is forced to split into two branches allowing a new $R_0$ condition that, at the single point $\lambda\approx 724$~nm and $\theta\approx 32^\circ$, is also an $R_0^\text{inv}$ condition since $s_\text{e}=1$ at this wavelength.

\section{Multipolar Origin of the Coupling Coefficient}
\label{app_multipoles}

In this appendix, we relate the coupling coefficient $d$, introduced in~\eqref{eq_ImWd}, to the multipole moments of the resonant mode responsible for the accidental BIC, which corresponds to the pole of $s_\text{o}$ at the complex frequency $\tilde{\omega}$. Since this pole belongs to the odd part of the response, the mode is odd under the mirror transformation $z\rightarrow -z$, so that it only radiates into, and is only excited by, the odd part of the illumination. Its multipole moments are those of the polarization currents of this field distribution inside the particle, normalized by the amplitude $a$ of the mode defined in~\eqref{eq_tcmt}. By reciprocity, $d$ is proportional to the overlap between the mode and the field of the odd-part illumination~\cite{fan2003}, which consists of two TM-polarized plane waves incident from both sides with the same magnetic field at the metasurface. Expanding this field around the center of the particle, located at $x=z=0$, each multipole moment of the mode couples to the field, or to its spatial derivatives, at this point. Specifically, $M_y$ couples to $H_y$, $P_z$ to $E_z$, the electric quadrupole $Q_{xz}^{\mathrm{(e)}}$ to the symmetric field gradient $\partial_xE_z+\partial_zE_x$, and the magnetic octupole $O_{yzz}^{\mathrm{(m)}}$ to the second derivative $\partial_z^2H_y$~\cite{achouri2022b}.

To obtain these quantities, consider a TM-polarized plane wave incident from the region $z<0$, with magnetic field $(H_0/2)\,e^{-j(k_xx+k_zz)}$, and its counterpart incident from the region $z>0$, with magnetic field $(H_0/2)\,e^{-j(k_xx-k_zz)}$. Since the electric field of each wave is orthogonal to its direction of propagation, $E_x=\pm\eta_0\cos\theta\,H_y$ for the waves propagating towards $\pm z$, while $E_z=-\eta_0\sin\theta\,H_y$ for both. The total field of this odd illumination is thus
\begin{subequations}
\label{eq_oddfield_total}
\begin{align}
H_y&=H_0\cos(k_zz)\,e^{-jk_xx},\\
E_x&=-j\eta_0\cos\theta\,H_0\sin(k_zz)\,e^{-jk_xx},\\
E_z&=-\eta_0\sin\theta\,H_0\cos(k_zz)\,e^{-jk_xx}.
\end{align}
\end{subequations}
Along $x$, the field carries the phase of the incident waves, so that each derivative with respect to $x$ brings a factor $-jk_x=-jk\sin\theta$. Along $z$, the two counter-propagating waves form a standing wave, in which $H_y$ and $E_z$ are maximum at $z=0$, while $E_x$ vanishes there, which is why this illumination only excites the odd part of the response. Each derivative with respect to $z$ thus brings a factor $k_z=k\cos\theta$. At the center of the particle, the dipoles therefore couple to $H_y=H_0$ and $E_z=-\eta_0\sin\theta\,H_0$. For the quadrupole, the two terms of the field gradient read
\begin{equation}
\partial_xE_z=j\eta_0k\sin^2\theta\,H_0,\qquad \partial_zE_x=-j\eta_0k\cos^2\theta\,H_0,
\end{equation}
where the first term results from the phase variation along $x$ and the second from the standing wave along $z$. Their sum is proportional to $\cos^2\theta-\sin^2\theta=\cos(2\theta)$, so that the quadrupolar coupling changes sign at $\theta=45^\circ$. Finally, since $H_y\propto\cos(k_zz)$, its second derivative is $\partial_z^2H_y=-k_z^2H_y$. In summary, we have
\begin{subequations}
\label{eq_oddfield}
\begin{align}
H_y&=H_0,\qquad E_z=-\eta_0\sin\theta\,H_0,\\
\partial_xE_z+\partial_zE_x&=-j\eta_0k\cos(2\theta)\,H_0,\\
\partial_z^2H_y&=-k^2\cos^2\theta\,H_0,
\end{align}
\end{subequations}
 Considering the dominant multipoles of Fig.~\ref{fig_multipoles}, the coupling coefficient is then of the form
\begin{equation}
\label{eq_d_multipoles}
\begin{split}
d(\theta)\propto\;& m_y-\eta_0\sin\theta\,p_z\\
&+c_Qk\cos(2\theta)\,q_{xz}+c_Ok^2\cos^2\theta\,o_{yzz},
\end{split}
\end{equation}
where $m_y$, $p_z$, $q_{xz}$ and $o_{yzz}$ are the magnetic dipole, electric dipole, electric quadrupole and magnetic octupole moments of the mode, normalized by its amplitude, and where $c_Q$ and $c_O$ are numerical factors that depend on the definition of the multipole moments. These factors also include the phase factors that relate each moment to the field it couples to, so that all the terms in~\eqref{eq_d_multipoles} have the same phase and $d$ is real, as expected for a lossless metasurface that is symmetric with respect to $z$.

The angular dependence of these moments follows from symmetry. Under the mirror transformation $x\rightarrow -x$, $M_y$, $Q_{xz}^{\mathrm{(e)}}$ and $O_{yzz}^{\mathrm{(m)}}$ change sign, while $P_z$ does not. Since the mode couples to the incident wave at normal incidence, it changes sign under this transformation, so that $m_y$, $q_{xz}$ and $o_{yzz}$ are even functions of $\theta$, while $p_z$ is an odd function of $\theta$ that vanishes at normal incidence, i.e., $p_z\approx p_1\sin\theta$. With $\cos(2\theta)=1-2\sin^2\theta$ and $\cos^2\theta=1-\sin^2\theta$, all the terms in~\eqref{eq_d_multipoles} are thus polynomials in $\sin^2\theta$. To lowest order, i.e., neglecting the angular variation of $m_y$, $q_{xz}$ and $o_{yzz}$, we obtain
\begin{equation}
\label{eq_d_ab}
d(\theta)\propto a+b\sin^2\theta,
\end{equation}
with
\begin{subequations}
\begin{align}
a&=m_y+c_Qk\,q_{xz}+c_Ok^2o_{yzz},\\
b&=-\left(\eta_0p_1+2c_Qk\,q_{xz}+c_Ok^2o_{yzz}\right),
\end{align}
\end{subequations}
which is the coupling coefficient of the passive model in~\eqref{eq_square}. The angular variation of the moments then leads to higher-order terms in $\sin^2\theta$, such as the envelope found in Sec.~\ref{sec_passive} for the silicon-ridge metasurface.

This result leads to three conclusions. First, since all the moments belong to the same mode, their contributions add in the coupling coefficient before being squared, so that the strength of the resonance, $W(\theta)=d^2$, is a perfect square. In contrast, independent responses, as in the diagonal dipolar model of~\eqref{eq_const}, add their strengths, which can only cancel each other with amplitudes of opposite signs. Second, an accidental BIC only requires one contribution in addition to $M_y$ whose coupling depends on the angle. This can be $P_z$, whose moment is proportional to $\sin\theta$, as in a dipolar but spatially dispersive model, or the quadrupole $Q_{xz}^{\mathrm{(e)}}$, whose coupling is proportional to $\cos(2\theta)$, as in a quadrupolar model with angle-independent susceptibilities. The octupole is thus not necessary for the BIC itself, but it contributes to $a$ and $b$ for the silicon-ridge metasurface, whose multipolar decomposition in Fig.~\ref{fig_multipoles} shows that $Q_{xz}^{\mathrm{(e)}}$ and $O_{yzz}^{\mathrm{(m)}}$ are comparable to $M_y$. Third, the coefficient $b$, which sets the angle of the BIC through $\sin^2\theta_\text{BIC}=-a/b$, does not only originate from $P_z$ but also from the quadrupole and the octupole. In a dipolar model, however, the only angle-dependent contribution is the term $k_x^2\chi_\text{ee}^{zz}$, so that all of $b$ must be attributed to $\chi_\text{ee}^{zz}$. This can lead to a large negative amplitude, $A_\text{ee}^{zz}=2ab$, even though $P_z$ itself is weak in Fig.~\ref{fig_multipoles}.

\section{Field Distribution of the Resonant Mode}
\label{app_fields}

\begin{figure}[t!]
\centering
\includegraphics[width=\linewidth]{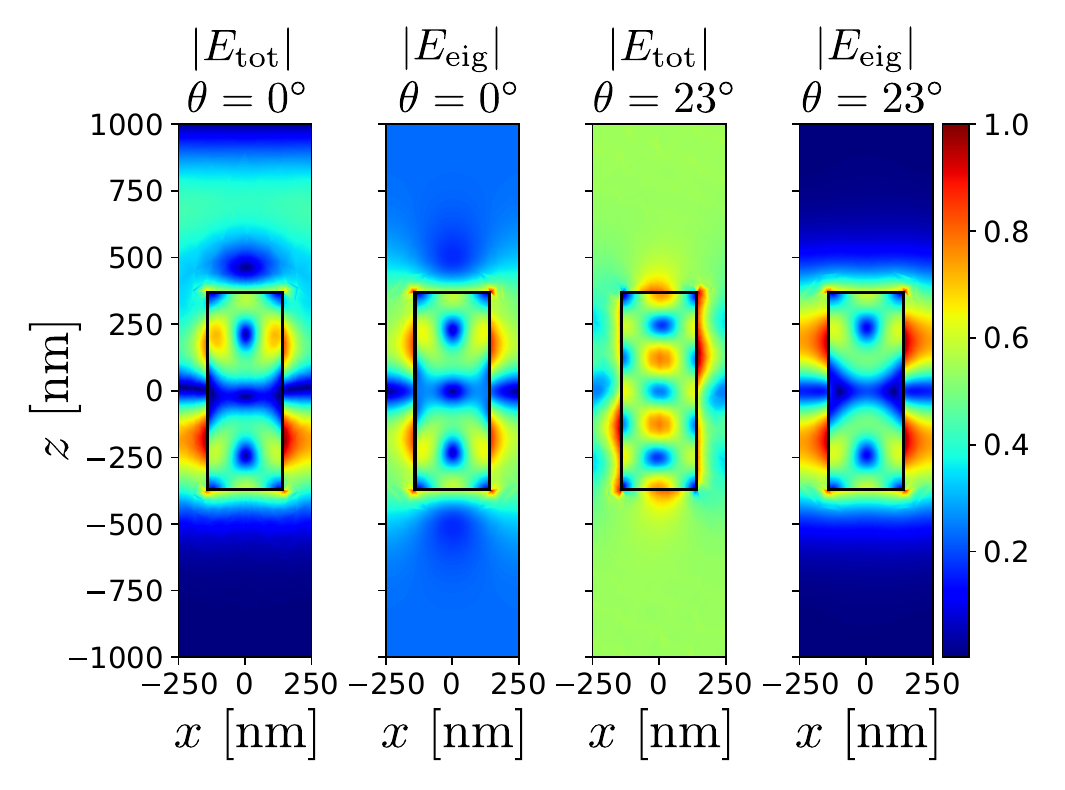}
\caption{Field distribution of the resonant mode of the silicon-ridge metasurface of Fig.~\ref{fig_BIC}. Norm of the total electric field, $E_{\mathrm{tot}}$ (incident plus scattered), and of the electric field of the resonant mode, $E_{\mathrm{eig}}$, for $\theta=0^\circ$ and at the BIC, $\theta=23^\circ$. The eigen-wavelengths and Q factor of this mode are shown in Fig.~\ref{fig_passive}.}
\label{fig_eigen}
\end{figure}

Figure~\ref{fig_eigen} compares the total electric field excited by the incident wave with the field of the resonant mode of the silicon-ridge metasurface, obtained from the eigenmode analysis of~\subsecref{sec_ridge}, for incidence angles of $\theta=0^\circ$ and $\theta=23^\circ$. At $\theta=0^\circ$, the total field strongly resembles the eigenmode field, indicating strong coupling between the incident wave and the resonant mode. Moreover, the field distribution inside the resonator clearly exhibits a multipolar pattern, which is consistent with the multipolar decomposition of Fig.~\ref{fig_multipoles}. In contrast, at $\theta=23^\circ$, which corresponds to the accidental BIC, the total field significantly differs from the eigenmode field, indicating that the incident wave no longer couples to the resonant mode. This total suppression of coupling is precisely what we expect from a bound state in the continuum.

\section{Metasurface with Omega-Type Bianisotropy}
\label{app_omega}

In this appendix, we consider the metasurface of~\secref{sec_model} when its mirror symmetry $z\rightarrow -z$ is broken by the omega-type bianisotropic susceptibilities $\chi_\text{em}^{xy}$ and $\chi_\text{me}^{yx}=-\chi_\text{em}^{xy}$, with the same normalization as in~\eqref{eq_biani}. We first derive the passivity condition~\eqref{eq_passive_cond} used in~\subsecref{sec_zbreak}, and then the corresponding scattering parameters and scattering zeros.

\subsection{Passivity Condition}
\label{app_omega_pass}

For a single susceptibility, Foster's reactance theorem, according to which the susceptibility of a lossless and passive response must increase with frequency~\cite{foster1924}, implies that the amplitude of a lossless Lorentzian resonance is positive, as discussed in~\subsecref{sec_accidental}. This ensures that the energy stored in the resonance is positive, so that the resonance can only give back the energy it has received from the incident wave and cannot act as a source of energy. When several susceptibilities share the same resonance and are coupled by bianisotropy, this must hold for any combination of the fields that excite it, here $E_x$ and $H_y$, whatever their relative amplitude and phase.

The resonant parts of the susceptibilities relate the fields $E_x$ and $H_y$ to the resonant polarizations through
\begin{equation}
\label{eq_res_matrix}
\begin{pmatrix} P_x\\ M_y \end{pmatrix}_\text{res}\propto\frac{1}{\Omega^2-\omega^2}
\begin{pmatrix} A_\text{ee}^{xx} & A_\text{em}^{xy}\\ (A_\text{em}^{xy})^* & W(\theta) \end{pmatrix}
\begin{pmatrix} E_x\\ H_y \end{pmatrix},
\end{equation}
with an appropriate normalization of the fields and polarizations, where the amplitude of $\chi_\text{me}^{yx}$ is the complex conjugate of that of $\chi_\text{em}^{xy}$, as required by reciprocity and losslessness, and where $W(\theta)$ is the strength of the resonance of $\chi_\text{mm}^{yy}$ and $\chi_\text{ee}^{zz}$. Note that, for a TM-polarized plane wave, $H_y$ is always accompanied by $E_z=-\eta_0\sin\theta\,H_y$, so that the excitation of the resonance through $P_z$ is already included in $W(\theta)$, via the term $k_x^2\chi_\text{ee}^{zz}$. For a single susceptibility, a field $E$ induces a polarization proportional to $AE$, and the energy stored in the resonance is proportional to the product of the field and of the polarization it induces, i.e., to $A|E|^2$, so that its sign is set by the amplitude $A$. Similarly, for an excitation by both $E_x$ and $H_y$, the energy stored in the resonance is obtained by multiplying the polarizations induced in~\eqref{eq_res_matrix} by the complex conjugates of the exciting fields, which shows that it is proportional to the effective amplitude
\begin{equation}
\label{eq_Atot}
W_\text{tot}=A_\text{ee}^{xx}|E_x|^2+W(\theta)|H_y|^2+2\,\text{Re}\{A_\text{em}^{xy}E_x^*H_y\},
\end{equation}
where the first two terms correspond to the excitation of the resonance by $E_x$ and $H_y$ alone, while the last term describes the interference between these two excitations through the bianisotropic coupling. Note that $E_x$ and $H_y$ can be chosen independently at the metasurface, since $E_x$ is set by the even part of the illumination and $H_y$ by its odd part, which can be combined with arbitrary amplitudes and phases. Passivity requires $W_\text{tot}\geq0$ for every excitation~\cite{bernland2011}, i.e., the positive amplitude implied by Foster's theorem must be ensured for any combination of $E_x$ and $H_y$. Considering $E_x=0$ or $H_y=0$ first shows that $A_\text{ee}^{xx}\geq0$ and $W(\theta)\geq0$. Moreover, by choosing the relative phase of $E_x$ and $H_y$, the last term in~\eqref{eq_Atot} can be made negative, down to $-2|A_\text{em}^{xy}||E_x||H_y|$. In this worst case,~\eqref{eq_Atot} can be rewritten as
\begin{equation}
\label{eq_Atot_worst}
\begin{split}
W_\text{tot}=&\left(\sqrt{A_\text{ee}^{xx}}\,|E_x|-\sqrt{W(\theta)}\,|H_y|\right)^2\\
&+2\left(\sqrt{A_\text{ee}^{xx}W(\theta)}-|A_\text{em}^{xy}|\right)|E_x||H_y|.
\end{split}
\end{equation}
The first term is never negative, but it vanishes for $|E_x|/|H_y|=\sqrt{W(\theta)/A_\text{ee}^{xx}}$. For this excitation, $W_\text{tot}$ reduces to the second term, which is non-negative only if $\sqrt{A_\text{ee}^{xx}W(\theta)}\geq|A_\text{em}^{xy}|$, which leads to~\eqref{eq_passive_cond}, i.e., the bianisotropic coupling can never exceed the geometric mean of the two direct terms. Conversely, if~\eqref{eq_passive_cond} is satisfied, both terms in~\eqref{eq_Atot_worst} are non-negative, so that $W_\text{tot}\geq0$ for every excitation.

Without a resonance of $\chi_\text{ee}^{xx}$, the condition~\eqref{eq_passive_cond} would thus be violated by any resonant bianisotropic susceptibility, so that some combinations of $E_x$ and $H_y$ would excite the resonance with a negative amplitude, and the model would no longer be passive. Moreover, the minimal choice in~\eqref{eq_zbreak}, $A_\text{ee}^{xx}=\alpha^2$ and $A_\text{em}^{xy}=j\alpha d$, satisfies~\eqref{eq_passive_cond} with equality, exactly as the perfect square in~\eqref{eq_square}, so that the resonance still corresponds to a single mode. A smaller bianisotropic amplitude would also be passive, but would correspond to an additional, independent resonance of $\chi_\text{ee}^{xx}$ at the same frequency, rather than to the asymmetry of a single mode.

\subsection{Scattering Parameters}
\label{app_omega_RT}

With bianisotropy and assuming reciprocity, the polarizations in~\eqref{eq_const} become
\begin{subequations}
\begin{align}
P_x&=\epsilon_0\chi_\text{ee}^{xx}E_{x,\text{av}}+\frac{1}{c}\chi_\text{em}^{xy}H_{y,\text{av}},\\
M_y&=\mu_0\chi_\text{mm}^{yy}H_{y,\text{av}}-\frac{1}{c}\chi_\text{em}^{xy}E_{x,\text{av}},\\
P_z&=\epsilon_0\chi_\text{ee}^{zz}E_{z,\text{av}}.
\end{align}
\end{subequations}
Substituting these expressions into~\eqref{eq_GSTC} and solving for the reflection and transmission coefficients of a TM-polarized plane wave incident from the region $z<0$ ($R_+$) or $z>0$ ($R_-$) leads to
\begin{subequations}
\label{eq_RT_omega}
\begin{align}
R_\pm&=\frac{2j\left(k_z^2\chi_\text{ee}^{xx}-Y\pm2kk_z\chi_\text{em}^{xy}\right)}{\Delta},\\
T&=\frac{k_z\left[4+\chi_\text{ee}^{xx}Y+k^2(\chi_\text{em}^{xy})^2\right]}{\Delta},
\end{align}
\end{subequations}
where $Y=k^2\chi_\text{mm}^{yy}+k_x^2\chi_\text{ee}^{zz}$ and
\begin{equation}
\label{eq_Delta_omega}
\Delta=(2+jk_z\chi_\text{ee}^{xx})(2k_z+jY)-k^2k_z(\chi_\text{em}^{xy})^2.
\end{equation}
For $\chi_\text{em}^{xy}=0$, these expressions naturally reduce to~\eqref{eq_RT}. The transmission coefficient is the same for both directions of incidence, as required by reciprocity, whereas the reflection coefficients differ.

Since the reflection coefficients now differ for the two directions of incidence, while reciprocity ensures that the transmission coefficient is the same, the scattering matrix of the metasurface reads
\begin{equation}
\label{eq_S_omega}
S=\begin{pmatrix} R_+ & T\\ T & R_-\end{pmatrix}.
\end{equation}
In contrast with~\eqref{eq_Smatrix}, the even and odd illuminations of Appendix~\ref{app_S}, i.e., $(1,1)$ and $(1,-1)$, are no longer eigenvectors of this matrix. To compare with the symmetric case, we nevertheless express $S$ in the basis formed by these two illuminations. This is done with the matrix
\begin{equation}
U=\frac{1}{\sqrt{2}}\begin{pmatrix} 1 & 1\\ 1 & -1\end{pmatrix},
\end{equation}
whose columns are the normalized odd and even illuminations, and which satisfies $U^\text{T}U=1$. In this basis, the scattering matrix becomes
\begin{equation}
\label{eq_UTSU}
\begin{split}
U^\text{T}SU&=\begin{pmatrix}
T+\dfrac{R_++R_-}{2} & \dfrac{R_+-R_-}{2}\\[2mm]
\dfrac{R_+-R_-}{2} & \dfrac{R_++R_-}{2}-T
\end{pmatrix}\\[1mm]
&=\begin{pmatrix}
s_\text{o} & s_\text{c}\\
s_\text{c} & -s_\text{e}
\end{pmatrix},
\end{split}
\end{equation}
where each diagonal entry is the projection of $S$ onto one of the illuminations, e.g., $\frac{1}{2}(1,1)\,S\,(1,1)^\text{T}$ for the odd one, while the off-diagonal entries describe the coupling between them. We thus have
\begin{subequations}
\begin{align}
s_\text{o}&=T+\frac{R_++R_-}{2},\\ 
s_\text{e}&=T-\frac{R_++R_-}{2},\\
s_\text{c}&=\frac{R_+-R_-}{2}.
\end{align}
\end{subequations}
This generalizes~\eqref{eq_seso} to
\begin{subequations}
\begin{align}
s_\text{o}&=\frac{(2+jk_z\chi_\text{ee}^{xx})(2k_z-jY)+k^2k_z(\chi_\text{em}^{xy})^2}{\Delta},\\
s_\text{e}&=\frac{(2-jk_z\chi_\text{ee}^{xx})(2k_z+jY)+k^2k_z(\chi_\text{em}^{xy})^2}{\Delta},\\
s_\text{c}&=\frac{4jkk_z\chi_\text{em}^{xy}}{\Delta}.
\end{align}
\end{subequations}
For a symmetric metasurface, $R_+=R_-$, so that $s_\text{c}=0$ and~\eqref{eq_UTSU} reduces to the diagonal form of~\eqref{eq_Smatrix}, with eigenvalues $s_\text{o}$ and $-s_\text{e}$. For $\chi_\text{em}^{xy}\neq0$, $s_\text{c}$ couples the even and odd parts of the response, and the eigenvalues of~\eqref{eq_UTSU} become $(R_++R_-)/2\pm\sqrt{T^2+s_\text{c}^2}$. Consequently, the poles of the scattering parameters are given by $\Delta=0$, i.e., $(2+jk_z\chi_\text{ee}^{xx})(2k_z+jY)=k^2k_z(\chi_\text{em}^{xy})^2$, where the right-hand side couples the resonances of the even and odd parts, whose poles are otherwise given by $2+jk_z\chi_\text{ee}^{xx}=0$ and $2k_z+jY=0$, respectively. The Q factor shown in Fig.~\ref{fig_zbreak}c is obtained by numerically solving $\Delta(\tilde\omega)=0$ for each angle of incidence, with $Q=\text{Re}\{\tilde\omega\}/(2\,\text{Im}\{\tilde\omega\})$.

\subsection{Scattering Zeros}
\label{app_omega_zeros}

From~\eqref{eq_RT_omega}, the $R_0$ condition for an incidence from either side reads
\begin{equation}
k_z^2\chi_\text{ee}^{xx}-Y\pm2kk_z\chi_\text{em}^{xy}=0.
\end{equation}
For a lossless and reciprocal metasurface, $\chi_\text{ee}^{xx}$ and $Y$ are real, while $\chi_\text{em}^{xy}$ is purely imaginary~\cite{achouri2021}, so that the real and imaginary parts of this condition must vanish separately, i.e., $k_z^2\chi_\text{ee}^{xx}=Y$ and $\chi_\text{em}^{xy}=0$. Reflection zeros are thus only possible if $\chi_\text{em}^{xy}$ vanishes, which is generally impossible as long as the structure remains broken for the mirror symmetry $z \rightarrow -z$. What actually happens instead is that the $R_0$ conditions becomes a reflection minima, whose depth is the same for both directions of incidence since $|R_+|=|R_-|$. In contrast, the $T_0$ condition,
\begin{equation}
4+\chi_\text{ee}^{xx}Y+k^2(\chi_\text{em}^{xy})^2=0,
\end{equation}
remains a real equation, since $(\chi_\text{em}^{xy})^2$ is real, so that the $T_0$ conditions still exist on the real frequency axis.

\bibliographystyle{ieeetr}

\bibliography{All.bib}

\end{document}